\documentstyle[epsf]{article}

\font\mften ptmr at 10 pt
\font\mftwe ptmr at 12 pt

\font\mftenb ptmb at 10 pt
\font\mftelb ptmb at 11 pt

\font\mfsixb ptmb at 16pt
\def\singlespace {\smallskipamount=3.75pt plus1pt minus1pt
                  \medskipamount=7.5pt plus2pt minus2pt
                  \bigskipamount=15pt plus4pt minus4pt
                  \normalbaselineskip=15pt plus0pt minus0pt
                  \normallineskip=1pt
                  \normallineskiplimit=0pt
                  \jot=3.75pt
                  {\def\smallskip {\vskip\smallskipamount}}
                  {\def\medskip   {\vskip\medskipamount}}
                  {\def\bigskip   {\vskip\bigskipamount}}
                  {\setbox\strutbox=\hbox{\vrule
                    height10.5pt depth4.5pt width 0pt}}
                  \parskip 7.5pt
                  \normalbaselines}
\def\middlespace {\smallskipamount=5.825pt plus1.5pt minus1.5pt
                  \medskipamount=11.25pt plus3pt minus3pt
                  \bigskipamount=22.5pt plus6pt minus6pt
                  \normalbaselineskip=22.5pt plus0pt minus0pt
                  \normallineskip=1pt
                  \normallineskiplimit=0pt
                  \jot=5.825pt
                  {\def\smallskip {\vskip\smallskipamount}}
                  {\def\medskip   {\vskip\medskipamount}}
                  {\def\bigskip   {\vskip\bigskipamount}}
                  {\setbox\strutbox=\hbox{\vrule
                    height15.75pt depth6.75pt width 0pt}}
                  \parskip 7.25pt
                  \normalbaselines}
\def\dblspc {\smallskipamount=7.5pt plus2pt minus2pt
                  \medskipamount=15pt plus4pt minus4pt
                  \bigskipamount=30pt plus8pt minus8pt
                  \normalbaselineskip=30pt plus0pt minus0pt
                  \normallineskip=2pt
                  \normallineskiplimit=0pt
                  \jot=7.5pt
                  {\def\smallskip {\vskip\smallskipamount}}
                  {\def\medskip   {\vskip\medskipamount}}
                  {\def\bigskip   {\vskip\bigskipamount}}
                  {\setbox\strutbox=\hbox{\vrule
                    height21.0pt depth9.0pt width 0pt}}
                  \parskip 15.0pt
                  \normalbaselines}

\def\be{\begin{equation}}

\def\j-{\J_-}

\def\ee{\end{equation}}

\def\bearr{\begin{eqnarray}}
\def\bearrs{\begin{eqnarray*}}
\def\eearr{\end{eqnarray}}
\def\eearrs{\end{eqnarray*}}
\def\barr{\begin{array}}
\def\earr{\end{array}}

\def\non\non{\nonumber}
\def\nn8{\nonumber\\[15pt]}
\def\l{\left}
\def\r{\right}

\def\f{\frac}

\def\dis{\displaystyle}

\begin{document}
\thispagestyle{empty}
\begin{center}
{\mfsixb Dynamic Predictions from Time Series Data - \\
An Artificial Neural Network Approach}\\[0.2in]
\mftwe
D. R. Kulkarni$^1$, A.S. Pandya$^2$ and J. C. Parikh$^1$\\[20pt]
\end{center}
\begin{tabular}{lp{5in}}
$^1$& Physical Research Laboratory, Navrangpura, Ahmedabad 380 009, India\\
$^2$& Dept. of Computer Sci. \& Engg., Florida Atlantic
University, Boca Raton, Florida, USA\\
\end{tabular}
\vspace {0.2in}
\begin{center}
\mftelb ABSTARCT\\[0.1in]
\end{center}
\singlespace
\mften
A hybrid approach, incorporating concepts of nonlinear dynamics 
in artificial neural networks (ANN), is proposed to model time
series generated by complex dynamic
systems.  
We introduce well known features used in the study of dynamic
systems - time delay $\tau$ and embedding dimension $d$ - for
ANN modelling of time series.  
These features provide a theoretical basis for selecting the
optimal size for the number of neurons in the input layer. 
The main outcome of the new
approach for such problems is that to a large extent it defines
the ANN architecture and leads to better predictions. 
 We illustrate
our method by considering computer generated periodic and
chaotic time series.
The ANN model  developed gave excellent quality of fit  for the training
and test sets as well as for iterative dynamic predictions for
future values of the two time series. Further, computer
experiments were conducted by introducing Gaussian noise of
various degrees in the two time series, to simulate real world
effects. We find rather surprising results 
that upto a limit  introduction of 
 noise leads to a
smaller network with good generalizing capability. \\[0.1in]
{\mftenb Keywords:~}\mften Nonlinear dynamics, Chaos, Artificial neural network, Time
Series Analysis.
\begin{center}
{\mftelb 1. \hspace{0.2cm} INTRODUCTION}\\
\end{center}
\mften
The study of nonlinear dynamics and chaotic systems has provided new
insights into the understanding of many complex physical systems and has
contributed significantly to the development of nonlinear time series
analysis$^{1,2,3}$. 
Formally, we expect that the complex dynamics is expressible in
the form of a set of coupled nonlinear differential equations or
discrete time maps involving all the system variables. The
dynamics of the system, in such ideal cases, may then be
investigated by solving the known equations. In most realistic
situations, however, such detailed information about all the
variables, their interactions and dynamics is not available.
Instead, for modelling the system we have time series data of
one or more significant system variables.  The task of
reconstructing the dynamics of the system from the time series
data is therefore of primary importance.  In this context, state
space
 methods of recovering the dynamics of the system in terms of 
the known time series have been proposed$^{1,2,4}$. The captured dynamics may then be used
to estimate invariants of dynamics as well as to make short-term predictions. 
Much of the success of neural networks is due to such
characteristics as parallel processing, nonlinear processing,
and nonparametric as well as distributed representation$^5$. The
prototypical use of neural nets is in {\it structural pattern
recognition}$^5$, where a collection of features - visual,
semantic or otherwise - is presented to a network and the
network must categorise the input feature pattern as belonging
to one or more classes.  In contrast, {\it temporal pattern
recognition} involves processing of patterns that evolve over
time.  The appropriate response at a particular point in time
depends not only on the current input, but potentially on all
previous inputs.  
Several researchers, including Miller et al$^7$, Weigend,
Hubernman and Rumelhart$^6$, Lapedes and Farber$^8$, Weigend
and Gershenfeld$^3$ etc. have discussed the paradigms for
system modelling using neural networks and time series data for
the actual system. 
 In this note, we, therefore, combine concepts and methods of 
nonlinear dynamics in the framework of ANN to construct a model
for the dynamics underlying
the time series. We also address the issue of how the
time-varying input sequence should be represented.  The nature
and quantity of information that must be presented to the
network is domain dependent.  \\[.2in]
In order to develop a model, it is 
necessary to subject the time series to rigorous pre-analysis phase. The 
pre-analysis characterizes the dynamic system 
and provides valuable information about its nature. In many cases, through
characterization, it is possible to find out whether the system is i) linear
or nonlinear, ii) deterministic or stochastic, iii) regular or chaotic. 
 For this purpose
various measures of the time series such as auto-correlation, power
spectrum, average mutual information, number of false nearest neighbours,
Lyapunov exponents and DVS (deterministic vs stochastic) plot
are evaluated. Kulkarni et al$^{9,10}$ discuss a similar
approach for analysing and modelling electroencephalograph (EEG)
time series data for capturing the underlying dynamics of the
system. 
It is important at this stage to understand the difference
between deterministic chaotic and stochastic (random) systems.
This is because the former can be modelled dynamically but the
latter cannot be - one has to use probabilistic notions.  A
chaotic system is one which exhibits extreme sensitivity to
initial conditions.  As a result a small error at time $t =
0$ grows exponentially with time $t$. Thus, even for such
completely deterministic systems the time evolution cannot be
predicted beyond a certain point as errors become large.  
Following the pre-analysis of the time
series data an appropriate ANN model is constructed 
if the system is dominantly deterministic and
predictions are made and tested.\\[.2in]
The organization of the paper is as follows:
In section 2, we describe the basis on which a model for the time series has
been developed. Section 3 outlines the implementation details of the
model in the framework of ANN. Various parameters
to characterize the system generating the time series are
described in section 4. In section 5, we illustrate the methodology
by applying  it to a periodic time series 
defined by the sine function whereas in section 6 the method is
used for a chaotic time series obtained by solving Lorenz
equations. 
In each case we consider a time series of about 1000 data
points.  This is in contrast to other studies in the
literature$^3$ where the time series has 10,000 points.  The
reason we choose a smaller set is because it is closer to real
world situation.
We also consider in sections 5 and 6 
effects of adding noise to these computer generated time series
since data of real system always has some noise.
 Finally, a summary and some conclusions are given in section 7. 
\vspace {0.1in}
\begin{center}
{\mftelb 2. \hspace{0.2cm} MODELLING THE TIME SERIES}\\
\end{center}
\mften
The modelling of complex dynamics from an observed time series
is based on the concept of state space  
reconstruction as advanced by  
 Packard et al$^1$ and Ruelle et al$^2$.
 They proposed a method called the method of delays to
reconstruct
the phase space using delay coordinate vectors derived from a scalar time series
$\l\{ X(t): t = 1,...N\r\}$ measured using time interval $\Delta$t.
A delay coordinate vector $\vec{X_i}$ is defined as\\
$$
\vec{X_i} = \l\{ X_i, \; X_{i-\tau},.....X_{i-(d-1)\tau}\r\}  \eqno(1)
$$
where $\tau$ is an appropriately chosen delay which is an integer multiple
of sampling time interval $\Delta$t, and $d$ is an integer
embedding dimension. If the solution 
to the equations lies on an attractor of dimension $d_A$, then
choosing an
integer $d > 2d_A$ is a sufficient condition for unfolding the attractor from the
scalar time series. The delay coordinate vectors or 
embedding vectors when evolved in time determine the essential geometric
and topological structures of the attractor of the complex system in the 
multi-dimensional space of its variables. Since the components of delay 
coordinate vectors are involved in modelling the system, it is vital that
these components be essentially independent. This has been generally ensured 
by choosing a
proper value of delay $\tau$. Another crucial parameter in this geometric
reconstruction of phase space is the embedding dimension $d$. Essentially 
it is the dimension of (Euclidean) phase space in which the
attractor is embedded or 
the minimum number of dynamic variables needed to model the
system. 
In view of these considerations, the time evolution of the
variable $X$
can be now given
in the form of a smooth map\\
$$
X_{i+T} = f \l\{ X_i, \; X_{i-\tau},......X_{i-(d-1)\tau} \r\}  \eqno(2) 
$$
where $T$ is the prediction step.\\
\vspace {0.1in}
\begin{center}
{\mftelb 3. \hspace{0.2cm} MODEL IMPLEMENTATION AND FORECASTING}\\
\end{center}
\mften
In order to implement the model in eq.(2), we need to know the crucial
parameters time delay $\tau$, the embedding dimension $d$ and the general
characteristics of the time series. To be specific, for a
deterministic system, the characteristics of
the series may indicate if the functional form of the model in
eq. (2) 
should be linear or nonlinear. 
 Since the exact
functional form in equation (2) is not known, a non-parametric model in
the framework of ANN$^3$ would in general be very useful.
Consequently, we employ the widely used 
multi-layer feed-forward network
architecture with backpropagation scheme to model the 
time series. The number of neurons in input and output layers are $N1 = d$
(embedding dimension) and $N2 = 1$ respectively. 
The number 
of neurons in the hidden layer  was varied to obtain an optimum network with least amount
of root mean square (rms) error. The transfer function used was hyperbolic
tangent for the hidden layer and linear one for the output layer. The
network is trained using the training set consisting of embedding vectors
and corresponding scalar prediction components (see eq. 2) derived from the
series. The last held-back values of the series, not included in the training
set, were used to compare the corresponding predicted values of the model.\\[.3in]
The future values were forecast using the iterated single step predictions.
The direct multistep predictions need different networks to be trained for
each value of $T$ and are thus very compute intensive. They had also been
reported$^{11}$ to be not as good as iterated single step predictions. In iterated
single step prediction, the previously predicted values were used to evaluate
future values. Thus the errors in the earlier predicted values get propagated
to the subsequently predicted values. The strategy thus seems to be a very
sensitive test of how accurately the dynamics has been captured. The measure
for the quality of prediction is given by the normalized mean square error (NMSE)
defined as\\
$$
\mbox{NMSE} = \f{\dis  \f{1}{N} \sum^N_{i=1} (\mbox{error})^2_i}
                 {\mbox{variance of } N \mbox{ (data points)}}  \nonumber
$$                       
where error(i) is the difference between i-th observed and predicted values.
The measure takes into account both the number of terms and range of two
sets to be compared. A value of $NMSE = 1$ corresponds to predicting the
unconditional mean.
\vspace {0.2in}
\begin{center}
{\mftelb 4. \hspace{0.2cm} PRE-ANALYSIS OF TIME SERIES}\\
\end{center}
\mften
In order to build an appropriate model, it is necessary to investigate
the characteristics of the series. As stated earlier, the characterization
gives both qualitative and quantitative information about the dynamical
system that generates the series. We discuss here some of the useful
quantities and their interpretation to understand the nature of the series.
Since 
some of these parameters have to be computed in the reconstructed phase space,
corresponding algorithms require a relatively large data set and are also sensitive
to the noise in the data which may lead to less accurate estimations of crucial 
parameters such as time delay $\tau$ and embedding dimension
$d$. We return to some of these points in Secs. 5 and 6 when we
add noise to the computer generated series. 
\begin{flushleft}
{\mftenb 4.1 \hspace{0.1cm} Auto-correlation function}\\
\end{flushleft}
\mften
A time series $\l\{ X(t): t = 1,2,.....,N\r\}$ is usually generated by measuring 
a dynamical scalar variable at some constant time interval $\Delta$t. The
behaviour of an auto-correlation function$^{4,12}$ (ACF) indicates how the successive
values are correlated with each other. An auto-correlation value $a(L)$ for
lag $L$ is defined as\\
$$
a(L) = \f{ \dis \sum^N_{i=1} \l( X (i+L) - \bar{X} \r) \l( X(i) - \bar{X}\r)}
         { \dis \sum^N_{i=1} \l( X(i) - \bar{X}\r)^2}  \eqno(4)
$$         
where $\bar{X}$ is the mean of the series. The ACF is a plot of
$a(L)$ versus $L$ for  
various values of lag $L$. If ACF falls to zero slowly, it implies that the
series is highly correlated and may have a dominant deterministic component
in the dynamics. However, if ACF falls to zero quickly, it may be that the
series is either stochastic or chaotic. In general the periodicity of the
series is well reflected in the periodicity of ACF.
\begin{flushleft}
{\mftenb 4.2 \hspace{0.1cm} Average mutual information}\\[1pt]
\end{flushleft}
\mften
A quantity called average mutual information$^4$ (AMI)
 is a probabilistic measure which is a generalization of auto-correlation
function to nonlinear domain. It is defined by the expression\\
$$I(L) = \sum^N_{i=1} P\l( X(i), X(i+L) \r) \log_2 \l[ 
         \f {\dis P\l(X(i), X(i+L) \r)} {\dis P\l(X(i)\r) P\l(X(i+L)\r)}
                                                   \r]  \eqno(5)
$$                                                   
where $P(X(k))$ is probability of measuring $X(k)$ and $P(X(k), X(k+L))$ is the
joint probability of measuring $X(k)$ and $X(k+L)$. The time delay $\tau$ used
in eq.(1) is generally chosen to be the value of $L$ at which the first
minimum in the  average mutual information (AMI) plot of $I(L)$
versus $L$ appears. Note that if there is no clear 
minimum in the plot, one takes $\tau$ to be that value of lag $L$ at which
$I(L) = I_{max}/5$. The time delay  $\tau$ can also be determined to be the lag
at which either the ACF crosses the zero line or it drops to 1/e. However,
the average mutual information method has an advantage that it considers 
all kinds of relations, not only the linear ones as in ACF. The best value of
$\tau$ is perhaps the one at which the phase space plot shows maximum scatter.
\begin{flushleft}
{\mftenb 4.3 \hspace{0.1cm} Power spectrum}\\
\end{flushleft}
\mften
The power spectrum is the fourier transform of the auto-correlation function.
The power spectrum is the plot of power versus frequency and clearly brings out 
the distinct periodicities in the form of sharp power peaks. However a broad
band spectrum may mean that the series is either stochastic or chaotic. In order
to differentiate it further, one needs to examine the fall of the power spectrum.
An exponential fall$^{13}$ with respect to frequency refers to chaotic series while
a power-law fall indicates that the spectrum is stochastic. It may be noted
that the sampling frequency and the length of the series need to be determined
on the basis of periodicities to be included in the model. The power spectrum
may also show the presence of stochastic or coloured noise in the form of low
power in the background.
\begin{flushleft}
{\mftenb 4.4 \hspace{0.1cm} Embedding dimension}\\
\end{flushleft}
\mften
As stated earlier, the embedding dimension determines the minimum number
of dynamic variables needed to model the system.
Several methods have been proposed to determine the embedding dimension d
from the time series data. The method$^4$ of false nearest neighbours is geometric
in construct. It rests on the fact that points in d dimension space may become 
neighbours if projected into a lower dimension.
Thus one calculates the number of false
nearest neighbours as a function of a variable dimension D. The value of D at 
which the number of false nearest neighbours goes to zero is taken to be the
embedding dimension. This method is intuitively appealing and computationally
quite robust. The plot of the number of false nearest neighbours
(FNN) as a function of
dimension D is very revealing. If the curve reaches zero and remains zero for higher
dimension, it implies that the dynamics is deterministic. However if the curve 
starts rising for higher dimension, it may imply that deterministic dynamics is 
to some extent diluted by stochastic or coloured noise. If it never falls to zero,
the dynamics may be interpreted to be predominantly stochastic. It may be stressed that
the data points in the series should not only be  sufficient in number but they
should also traverse the major part of the attractor. The goodness of time series
prediction largely depends on the fact that dynamics is dominantly deterministic.
\begin{flushleft}
{\mftenb 4.5 \hspace{0.1cm} Lyapunov exponents}\\
\end{flushleft}
\mften  
Chaos in deterministic system implies a sensitive dependence on initial condition.
Based on this property,
Lyapunov exponents provide a diagnostic tool to determine whether or not a system
is chaotic. The basic idea is that for chaotic systems if two
trajectories start close to one another 
in phase space, they will move exponentially away from each other for small times
on the average. Thus if $d_0$ is a measure of the initial distance between the two 
starting points, at a small but later time the distance is \\
$$
d_t = d_0 \exp \l( \lambda t \r)  \eqno(6)
$$      
where $\lambda$ is called a Lyapunov exponent. Lyapunov exponent is defined for each
dimension of phase space and thus what we get is Lyapunov spectrum.
There are many algorithms to determine the values of exponents from the time series.
We have used algorithm by Zeng et al$^{14}$ which can be used for short time series to
evaluate the Lyapunov exponent spectrum.
 If the
largest Lyapunov exponent is greater than zero, then the system is chaotic. Otherwise
it is considered as regular motion.
\begin{flushleft}
{\mftenb 4.6 \hspace{.1cm} DVS Plot}\\
\end{flushleft}
\mften
The DVS plot provides very broad qualitative understanding of dynamics of time series.
The DVS algorithm$^{15}$ attempts to fit an auto-regressive linear model of the form\\
$$
X_{(i+T)} = f \l( \vec{X_i} \r)  \eqno(7)
$$
where $f$ is a linear function and $\vec{X_i}$ is the i-th
embedding vector as defined in section 2.
In order to obtain the DVS plot, we fit a family of linear models varying the size of
neighbourhood in a reconstructed phase space. The short-term predictive accuracy
of the models is estimated as out-of-sample error on a held-back portion of the
time series. We then plot the average out-of-sample errors as a function of the size of 
the neighbourhood to obtain what is called a DVS plot. If the underlying dynamics is 
indeed chaotic and low dimensional, the models with small neighbourhood  should give
more accurate forecasts than those with large neighbourhood. This is because the 
nonlinearity in small neighbourhoods can be very well approximated by linear models.
However if the underlying dynamics is stochastic or chaotic with high dimension, then
very local models (in small neighbourhoods) will give less accurate forecasts
as they tend to fit the noise in addition to the signal compared to global models with
large neighbourhood. One can also vary the embedding dimension systematically to find
which value of embedding dimension gives smallest error. This may be an alternative way 
of determining the value of embedding dimension. We have obtained DVS plots using local
linear approximation as described here. It may be possible to obtain the DVS plots
using nonlinear models using radial basis functions or neural networks.\\[.2in]
\begin{center}
{\mftelb 5. \hspace{0.2cm}MODELLING A PERIODIC SYSTEM}\\
\end{center}
\mften

As a simple illustration we first 
model the dynamics
of a sine series, using sampled data which provided the time
series. The system output, $y$ was given as follows: 
\[
y = A * \sin ( 2\pi f + \theta )       
\]
 where  $A$ = 1.0, $f$ = 2.0  and   $\theta$ = 0   were the
chosen as parameter values.\\[.2in] 
The sampling frequency for generating the time series data was
chosen to be  256 Hz and thus the time step $\Delta T$ was taken as 1/256.
The network was trained with the data containing about 7 full
cycles which provided 950 data points, while the next 300 data
points had to be predicted during the testing phase. 
The time delay ($\tau$) was
determined from the evaluation of average mutual information and
found to have the value 4.
Based on the value of   $\tau$ = 4 , the embedding dimension, $d$,  value was
found to be  3. \\[.2in]
	A multilayer perceptron with 3$\times$7$\times$1 configuration was
trained using the input-output vectors obtained with the above parameters.  The network
converged to the NMSE value of 0.00009 and the fit of the
trained network to the observed 950 data values was nearly
perfect. Using the trained network single step predictions were
made for the next 300 values in the time series and again the
fit was nearly perfect. However in order to determine whether
the neural network has really extracted the underlying dynamics
of the system (i.e. frequency, amplitude, phase etc. in case of
this system) it is essential to test its performance for
iterated single step predictions. Fig. 1 shows the
iterated single step prediction values which demonstrate a
complete failure in terms of all the system characteristics
(i.e. frequency, amplitude etc.). The NMSE for the predicted
values for the first 50 points is  208, which is extremely high
and implies that the network is not able to generalize.\\[.2in]
	As a result, the number of neurons in the hidden layers
were increased and the network was trained for various
configurations. It was determined that a configuration of 3x20x1
was required for training the network for convergence to a low
NMSE of 0.000003. Fig. 2 shows the observed values 
and the iterated single step
predictions using this trained network
for the next 300 points in the sine series. It is clearly
evident that the network model has captured the characteristics
of the complex system very well in terms of the frequency,
amplitude, phase etc. In this case for the predicted 300 points
the NMSE was 0.003. \\[.2in]
For time series predictions often recurrent networks
perform better than the traditional multilayer feed forward
architecture. The network was then trained using the
Backpropagation Through Time (BTT) method$^{16}$ using a
correction for various step sizes. It was seen that a much
smaller network architecture (3x6x1) was required in order to
model the system in this case.  See Fig. 3 for the plots of
observed data versus iterated single step prediction values for the next 300
data points generated at the same sampling rate. It can be seen
that the process of feedback adds the necessary non-linearity,
therefore a smaller network is able to perform the same task. \\[.2in]
	At this stage, a new data set was generated by
superimposing noise with gaussian distribution on the original
sine series data set. Gaussian noise with mean = 0 and the
standard deviation equal to some percentage of the standard
deviation of the signal into which it is to be superimposed was
generated. More precisely, the new series is$^{17,18}$
\[
Y_i = X_i + \alpha Z_i \; \; (i = 1,...N)
\]
where $\l\{X_i\r\}$ is the original series and the set $\l\{ Z_i
\r\}$ denotes $N$ randomly chosen Gaussian variates with mean
$\bar{Z} = 0$ and variance $\sigma^2_Z = 1.0$.  The parameters
$\alpha \equiv k \f{\sigma_X}{\sigma_Z} =  k\sigma_X$ where
$\sigma^2_X$ is the variance of the time series $\l\{X_i\r\}$.
Further, $k$ in per cent denotes the amount of noise we add to
the original pure (noise free) time series $\l\{ X_i \r\}$.\\[.2in] 
The signal to noise ratio (SNR) was varied such
that the degree of noise $k$ was initially 5\% and then it was
increased in steps of 5\% to a total of 70\%. The results of the
fit and predictions for 5, 15 and 28\% noise were excellent.
Here we report the results for 28\% noise. For the noisy data set
the time delay ($\tau$) as well as the embedding dimension ($d$)
increased compared with the pure sine series data. Fig. 4 shows
the plot of time delay and the embedding dimension. It  was
observed  that the dip in the value for $\tau$ was not as prominent
as seen in case of the pure data. In case of the $d$ values the
number of false neighbours remains zero throughout, indicating
the dominantly deterministic nature of  the system in the presence of
noise upto 40\%. 
As the noise increased beyond 40\% the plot of number of false
nearest neighbour versus variable dimension shows a different
behaviour.  Initially the number of false nearest neighbours
decreases with $D$ and then increases again.  This is shown in
Fig. 5. Such behaviour is indicative of stochastic nature of the
system.  As mentioned earlier statistical (not ANN) techniques
are required to model the system in these cases.  
 Thus one can expect the neural network to extract the
trend and model the system for noise upto 40\%
since the signals are not random even
though there is noise.\\[.2in]
While modelling the time series with noise it was observed that
for noise upto 15\% the network had fewer parameters compared to
the pure case.  By the time the noise was 28\% the network
size had again become large.  This is shown in Table 1.
Based on the plots of Fig. 4, we obtained $\tau = 8$ and $d =
5$ for 28\% noise. A network with at least 10 hidden
 neurons were required for convergence at a lower NMSE
value of  0.05. Increasing the number of hidden layer neurons
beyond this resulted in overfitting, such that the network
converged to a lower NMSE value but the characteristic features
like the frequency, amplitude etc. were not correctly extracted.
Fig. 6 shows the plot of noisy data set with 28\% noise and the
fit obtained by training the neural network. \\[.2in]
	Figures 7a and 7b show the next 300 actual data points for this
series along with the iterated single step
prediction values. In this case the first input vector used for
the iterated single step predictions was the noisy data vector. The NMSE of
the predicted values with respect to noisy data was found to be 0.066.
We observe that the plot for the iterated single step predictions contains
jitters which subside as you predict further ahead in time.  Figure 7a
indicates the noisy input data values along with predicted values. It can be
seen that the trend is extracted quite nicely in the presence of noisy data.
Next we have compared these predicted values with the
noise-free sine series values. The corresponding NMSE was 0.024 which is
smaller than the value of 0.066 obtained with noisy data. This
implies that the network has really captured the dynamics and has good
generalizing capability. This was further reinforced when the iterated
single
step predictions were obtained using the noise-free sine series values as
initial input values. The NMSE
of these predictions with respect to the corresponding values of noise-free
sine series was found to be 0.0036. Since the starting points have
no error, the network could generate the original noise-free sine series
to a great deal of accuracy.
Consequently the plot in fig. 7b which shows the actual pure sine series
values along with the predicted values  is very smooth.
 It can be
seen that the neural network models the system perfectly even
after being trained using noisy data and the predictions are
even better than the network trained with pure sine data (Fig.
2), even though the NMSE value at convergence is higher (0.05
vs. 0.000003) in this case.
It may also be noted that the number of parameters in this network is
smaller (71) than those in the network trained with pure sine seris (101)
as seen in Table 1.\\[0.2in]
\begin{center}
{\mftelb 6. \hspace{0.2cm} MODELLING A CHAOTIC TIME SERIES}\\
\end{center}
\mften
The systematic methodology of time series prediction is next
applied to a computer generated time series of variable X obtained
by solving Lorenz equations given below.
\begin{eqnarray}
\dot{X}&  =  &  \sigma(Y - X) \nonumber \\
\dot{Y}&  =  &  \rho X - Y -XZ \nonumber \\
\dot{Z}&  =  &  -\beta Z + XY  \nonumber
\end{eqnarray}
The parameters used in the solution of these equations are $\sigma = 16.0$, $\rho = 45.92$
and $\beta = 4.0$. The first 5000 points in the solution have been omitted to get rid of 
transient effects. Subsequently the series of X variable has been calculated by 
integrating the equations using the time step of 0.001 second for better 
accuracy. However the actual time series used for present analysis has  obtained 
using the sampling interval $\Delta t = 0.05$ by selecting every 50-th value from
the calculated series. Thus we obtained a Lorenz's time series of 1050 values.
Although we could have taken much longer time series with obvious associated
advantages, we preferred the size of 1050 values which is a typical size generally
encounterd in scientific time series analysis. It may also be noted that the 
Lorenz series is a well known example of chaotic series.\\[0.3in]
In the pre-analysis phase we obtained ACF plot (fig. 8a), AMI plot (fig.
8b), FNN plot (fig. 9),
power spectrum (fig. 10) and DVS plot (fig. 11). We also calculated the
largest Lyapunov exponent.
The ACF plot showed that the values in the series are quite correlated and
auto-correlation value goes to zero only at $lag = 25$. If we had to  determine
the time delay from ACF, we would obtain $\tau = 4$ where the ACF approximately 
assumes the value of 1/e. However, as discussed before, we preferred to deletrmine
the value of time delay $\tau$ based on AMI plot. Thus we obtained the value of
time delay $\tau$ to be 2, where the AMI plot (see fig. 8b) had a distinct
minimum.\\[0.2in]
The value $\tau$ is the input to obtain FNN plot which enables us to get the value
of embedding dimension. Using $\tau = 2$ we obtained the FNN
plot (Fig. 9). It was
clear from the plot that the value of embedding dimension for the present series is 3
where the plot reached its zero value. It is gratifying to know that the Lorenz series
of finite size could yield the correct value of the dimension of the state space.
Since in linear time series analysis $\tau$ is generally assumed to be always one,
we also obtained the value of embedding dimension using $\tau = 1$. The fact that
it happened to be 3 seems to be just coincidence. In principal
it can be different.\\[0.2in] 
Being a chaotic series, the power spectrum of the present series was worth
examining. We obtained (fig. 10) the plot  of log(power) versus frequency.
It was satisfying to
find that the plot was just a straight line confirming that power spectrum of
chaotic signal has an exponential power fall. The conclusion that the present
series is chaotic was further supported by the evaluation of the largest Lyapunov
exponent which was found to be 1.48.\\[0.2in]
The further probe of the series by using DVS plot was quite useful. The DVS plot
(fig. 11) was obtained using the embedding dimension $d = 3$. We found that
for $d=3$,
the RMS error was minimum.  The curve also showed that RMS error went on
increasing as we enlarged the neighbourhood. These two observations led us to
broad qualitative understanding that the present series belonged to low-dimensional,
nonlinear and deterministic system.\\[0.2in]
Next we tried to model the series using the information obtained from pre-analysis
We developed following two models, using eq. (2). 
\begin{enumerate}
\mften
\item Using $\tau = 1$ and $d = 3$
\item Using $\tau = 2$ and $d = 3$
\end{enumerate}
The models were implemented using multi-layer feed-forward ANN with a single hidden
layer. The first 950 values of the series were used to train the
network and 100 additional 
values were predicted using the trained network. The predictions were made
using both  single-step prediction (SSP) 
and iterated single-step prediction (ISSP) scheme as described earlier.
 The parameters and results of the models are given in Table 2.
Figure 12  shows the plot of 50 predicted values of the series obtained
using model 1. Figure 13 shows the plot of 50 predicted values of the
sine series obtained using model 2.\\[0.2in]
It can be seen from table 2 that in both the models the SSP values are in
excellent agreement with 
the observed values. So far as the ISSP values are concerned, the first model could
not capture the dynamics properly despite its good quality fit to training set
and SSP values. Thus a good fit and equally good SSP values may not gaurantee  good
ISSP values. In fact we found that the NMSE value for the first 30 values in the
first model is 0.085 which is substantially larger than the corresponding value
(0.0108) in the second model. In contrast the second model could give the ISSP
values which are in good agreement with the corresponding observed 50 values.
This implies that the considerations of nonlinear dynamics in modelling the series
are quite crucial. We also found that the predictability of the
model varied with the initial point chosen - i.e. whether it was
949, 950 or 951.  Further, in all the cases the breakdown of the
model occurred suddenly - as seen in Fig. 13.  These features
are due to the chaotic nature of the system as they indicate
sensitivity to initial conditions.\\[0.2in]
We next introduce noise in the Lorenz series.  It has been found
that, unlike the regular sine series, the modelling and
prediction results in the case of chaotic Lorenz series are very
sensitive to noise.  While the sine series could be successfully
modelled upto k=40\% of noise, the Lorenz series could tolerate
noise only upto k=0.4\%.  Figure 14 shows the effect of
increasing the noise from k= 0.4\% to k = 1.6\% on embedding
dimension.  Note that for the noise level $>$ k= 0.4\%, the FNN
plot shows a rise with increase in the varying dimension $D$.
Even for very small noise level of k = 0.4\%, the embedding
dimension value of Lorenz's series changed to 4 from 3.  This
implies that the system becomes predominantly stochastic even at
low level of noise.  As before for lower values of noise we do
observe a reduction in the number of parameters required to model
the system as we go from pure (noise free) limit to noise of
k = 0.4\%.  For larger values of k, however,  we cannot model the system
using ANN as it is predominantly stochastic as indicated by the
FNN plot (Fig. 14).\\[0.2in]
For gaussian noise with $k = 0.2\%$
 superimposed on the pure data set (1000 points) generated
earlier, the $\tau$ and $d$ values were calculated as $\tau = 2$
and $d = 3$.  A neural network with configuration
$3\times7\times1$ using the noisy data set
has converged to an NMSE value of
0.002.  Figure 15 shows the plot of actual data
and the
iterated single step prediction values  by the trained
network for the next 50 points sampled at the same rate.  Again
it was found that the fit was almost as good as in the case of the network
trained with the pure data with $NMSE = 0.095$ for the first 50 predicted
values.\\[0.2in]
\begin{center}
{\mftelb 7. \hspace{0.2cm} SUMMARY AND CONCLUSIONS}\\
\end{center}
\mften
In this paper, we have modelled computer generated periodic and
chaotic time series using ANN techniques. Before modelling the
system we studied the characteristics of the underlying dynamics
of the time series using the tools of nonlinear dynamics. The
analysis of time series data enabled us to partly determine the
architecture of the network. More precisely, we argue that the
number of neurons in the input layer ought to be equal to the
embedding dimension $d$. Our results for iterated dynamic
predictions establish the importance of the parameter $\tau$
(time lag).  We have also examined the effect of adding white
noise to the two model time series.  It is explicitly
demonstrated that upto a limit noise helps in the sense that the
network is smaller and the predictions better and robust.
However, as the noise level increases, the determinstic
component gets diluted and modelling becomes difficult.  This
can be very well seen by the study of FNN plot in which the
false nearest neighbours increases considerably for higher
dimension. Our study has also revealed that the noise tolerance
level is much higher (about 40\%) in the regular sine series
than in the chaotic Lorenz series (about 0.4\%). Finally, in our
view, this study establishes for the first time, a \underline{systematic
way}  of constructing non-parametric models for time series data.
\newpage
\begin{center}
{\mftelb \hspace{0.2cm} REFERENCES}\\
\end{center}
\begin{enumerate}
\mften
\item N.H. Packard, J.P. Crutchfield, J.D. Farmer, and R.S. Shaw,
Geometry From Time Series, {\it Phys. Rev. Lett.} {\bf 45,}  pp. 712-716, 1980.
\item D. Ruelle and J.P. Eckmann, Ergodic Theory of Chaos and Strange Attractors,
{\it Rev. Mod. Phys.} {\bf 57,} pp. 617-656, 1985.
\item A.S. Weigend and N.S. Gershenfeld (editors), {\it Time Series
Prediction,} Addison Wesley (Reading, Massachusetts), 1994.
\item H.D.I. Abarbanel, R. Brown, J.J. Sidorowich and L.Sh.
Tsimiring, The Analysis of Observed Chaotic Data in Physical
systems, {\it Rev. Mod. Phys.} {\bf 65,}, pp. 1331-1392, 1993.
\item A.S. Pandya and R.B. Macy, {\it Pattern Recognition with
Neural Networks in C$^{++}$}, CRC Press, Boca Raton, and IEEE
Press, 1995.
\item W.T. Miller, R. Stutlon and P.J. Werbos (editors), {\it
Neural Networks for Control}, MIT Press, Cambridge, 1992.
\item A.S. Wiegend, B.A. Hubberman and D.E. Rumelhart, {\it
Predicting the Future : A Connectionist Approach}, International
Journal of Neural Systems, {\bf 1}, pp. 193-209, 1990.
\item A. Lepedes and R. Farber, {\it Nonlinear Signal Processing
Using Neural Networks}, Tech. Rep. No. LA-UR-87-2662, Los Alamos
Lab., Los Alamos, NM, 1987.
\item D.R. Kulkarni, J.C. Parikh and R. Pratap, Simulation of
Characteristics and ANN Modelling of EEG Time Series, {\it
Phys. Rev.} {\bf E} (in press).
\item D.R. Kulkarni and J.C. Parikh, {\it Study of EEG Signals
using Nonlinear Dynamics and ANN}, Proc. of In. Conf. on
Cognitive Sciences (ICCS), New Delhi, 1996.
\item J.D. Farmer and J.J. Sidorowich, Exploiting Chaos to Predict the Future
and Reduce Noise, in {\it Evolution, Learning, and Cognition}, edited by
Y.C. Lee, World Scientific, Singapore, 1988.
\item C. Chatfield, {\it The Analysis of Time Series}, Chapman and Hall, 1980.
\item U. Frisch and R. Morf, Intemittancy in non-linear dynamics and singularities
at complex times, {\it Phys. Rev.} {\bf 23}, pp. 2673, 1981.
\item X. Zeng, R. Eykholt and R.A. Pieke,
Estimating the Lyapunov-Expoenent Spectrum from Short Time Series of Low Precision,
{\it Phys. Rev. Lett.} {\bf 66}, pp. 3229-3232, 1991.
\item M. Casdagli and A.S. Weigend, Exploring the Continuum between Deterministic and
Stochastic modeling, in {\it Time Series Prediction} (see ref. 3).
\item P.J. Werbos, {\it Backpropagation through time : What it does
and and know to do it}, Proc. of IEEE, {\bf 8}, pp. 1550-1560,
1990. 
\item Ramzan, G., Nonlinear Predictionof Noisy Time Series
with Feed-forward Network, {\it Phys. Letts.}, {\bf A187}, pp.
397-403, 1994.
\item A.S. Pandya, D.R. Kulkarni, and J.C. Parikh, {\it Study of
Time Series Prediction under Noisy Environment}, Proc. of SPIE
Conf. on Applications and Science of Artificial Neural Networks
III, 1997.
\end{enumerate}
\newpage
\begin{center}
\bf{Table 1}\\
Network Configuration for Sine Series Prediction\\[0.15in]
\begin{tabular}{|c|c|c|}
\hline\hline
Data&Configuration&No. of Parameters\\
\hline\hline
Pure Sine Series&3-20-1&101\\
\hline
Sine Series + 5\% Noise&4-4-1&25\\
\hline
Sine Series + 15\% Noise&3-6-1&26\\
\hline
Sine Series + 28\% Noise&5-10-1&71\\
\hline\hline
\end{tabular}
\end{center}
\vspace*{0.5in}
\begin{center}
\bf{Table 2}\\
Configuration and Errors in Two Models of Lorenz Series\\[0.15in]
\begin{tabular}{|c|c|c|c|c|c|c|} \hline\hline
\multicolumn{1}{|c|}{Model} &
\multicolumn{1}{c|}{Network} & NMSE & NMSE & NMSE &
\multicolumn{1}{|c|}{NMSE}&\multicolumn{1}{|c|}{NMSE}\\
&
\multicolumn{1}{c|}{configuration} & (Training) & (SSP-50) & (ISSP-30) &
\multicolumn{1}{|c|}{(ISSP-50)}&\multicolumn{1}{|c|}{(ISSP-100)} \\ \hline \hline
1 & (3-7-1) & 0.0002 & 0.00008 & 0.0850 & 1.4580 & 1.8640\\
2 & (3-7-1) & 0.0003 & 0.00019 & 0.0108 & 0.0476 & 0.1570\\
\hline
\end{tabular}
\end{center}
\newpage
\begin{center}
{\bf Figure Captions}\\
\end{center}
\begin{itemize}
\item[Fig. 1] Actual (dashed line) and iterated single step
predictions (solid line) of the sine series using a multi-layer
perceptron with configuration (3x7x1).
\item[Fig. 2] Actual (dashed line) and iterated single step
predictions (solid line) of next 300 (out-of-sample) values for
the sine series.  The network configuration was (3x20x1).
\item[Fig. 3] Actual (dashed line) and iterated single step
predictions (solid line) of next 300 values of the sine series
using recurrent network with a configuration (3x6x1).
\item[Fig. 4] a) A plot of average mutual information versus lag
for the sine series with noise for SNR at k = 28\%.\\
b) A plot of false nearest neighbours versus dimension for the
data in (4a).
\item[Fig. 5] A plot of false nearest neighbours versus
dimension for the sine series with noise for SNR at k=56\%
(dashed line) and k=70\% (solid line).
\item[Fig. 6] Actual (dashed line) and fitted values (solid
line) of the sine series with noise for SNR at k = 28\%.
\item[Fig. 7] The iterated single step prediction (solid line)
values for the next 300 values of the sine series with noise for
SNR at k = 28\% using the network configuration (5x10x1) along with
\begin{itemize}
\item[a)] The actual values (dashed line) of the data with noise
and noisy data as a starting point for the predictions.
\item[b)] The actual values (dashed line) of the pure sine
series and pure data as a starting point for the predictions.
\end{itemize}
\item[Fig. 8] a) The auto-correlation function for the pure
Lorenz series.\\
b) The plot of average mutual information versus lag for the
pure Lorenz series.
\item[Fig. 9] The plot of false nearest neighbours versus dimension
with $\tau = 2$ for the pure Lorenz series.
\item[Fig. 10] The plot of power spectrum for the pure Lorenz
series.
\item[Fig. 11] The DVS plot for the pure Lorenz series, with
$\tau = 2$ and $d = 3$.
\item[Fig. 12] The actual (dashed line),
 and iterated single step prediction (solid line) values
fo the next 50 (out-of-sample) values of the Lorenz series
using model 1.
\item[Fig. 13] The actual (dashed line),
 and iterated single step prediction (solid line)
values of the next 50 (out-of-sample) values of the Lorenz
series using model 2. 
\item[Fig. 14] The plot of false nearest neighbours versus
dimension for the noisy Lorenz series with\\
a) SNR at k = 0.4\% (dashed line)
b) SNR at k = 0.8\% (dotted line)
c) SNR at k = 1.6\% (solid line)
\item[Fig. 15] The actual (dashed line),
 and iterated single step prediction (solid line)
values of the next 50 (out-of-sample) values of the noisy
Lorenz series with SNR ar k = 0.2\%. The network configuration
was (3x7x1). 
\end{itemize}
\newpage
\begin{center}
\leavevmode
\epsfxsize 6in
\epsfbox{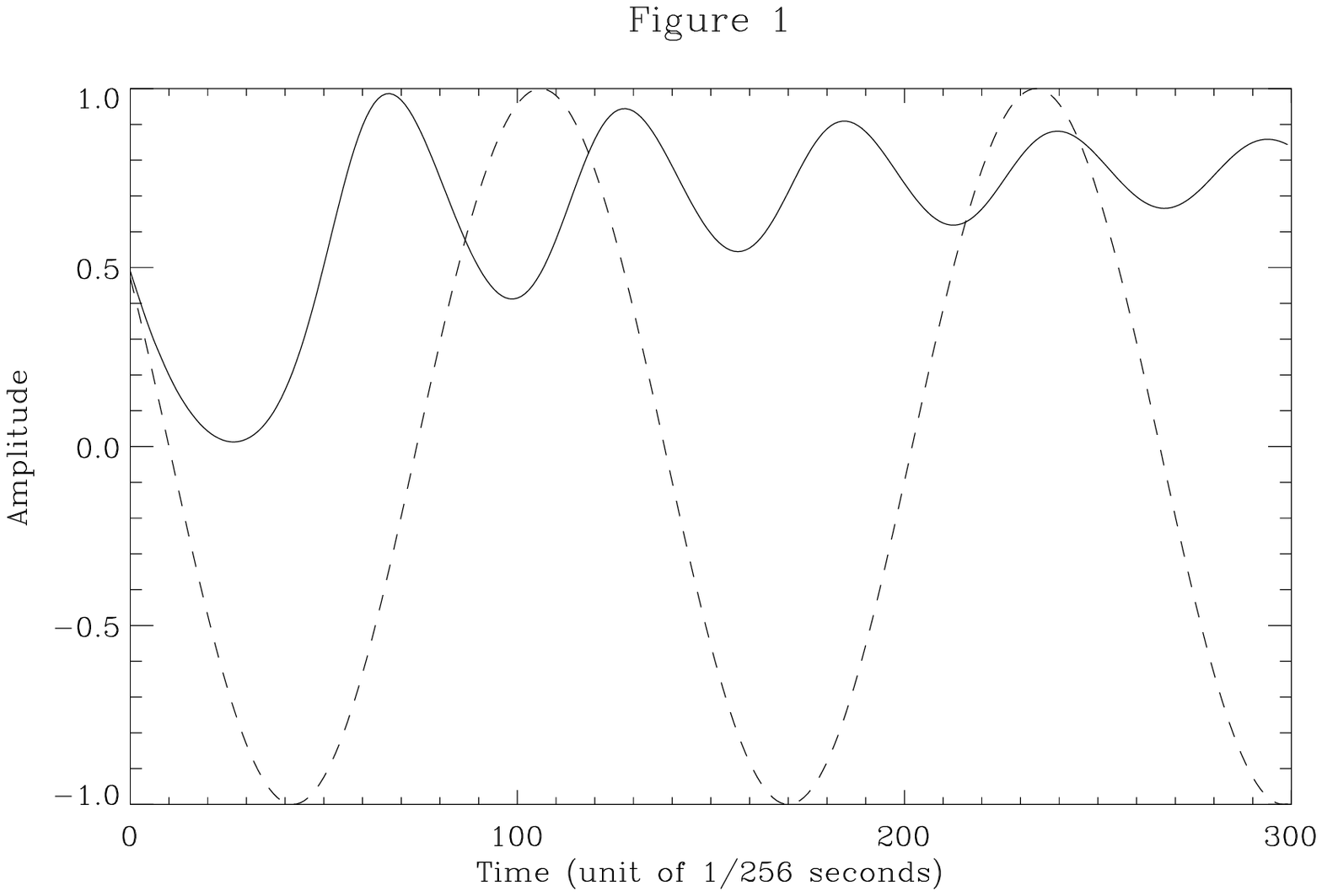}
\end{center}
\newpage
\begin{center}
\leavevmode
\epsfxsize 6in
\epsfbox{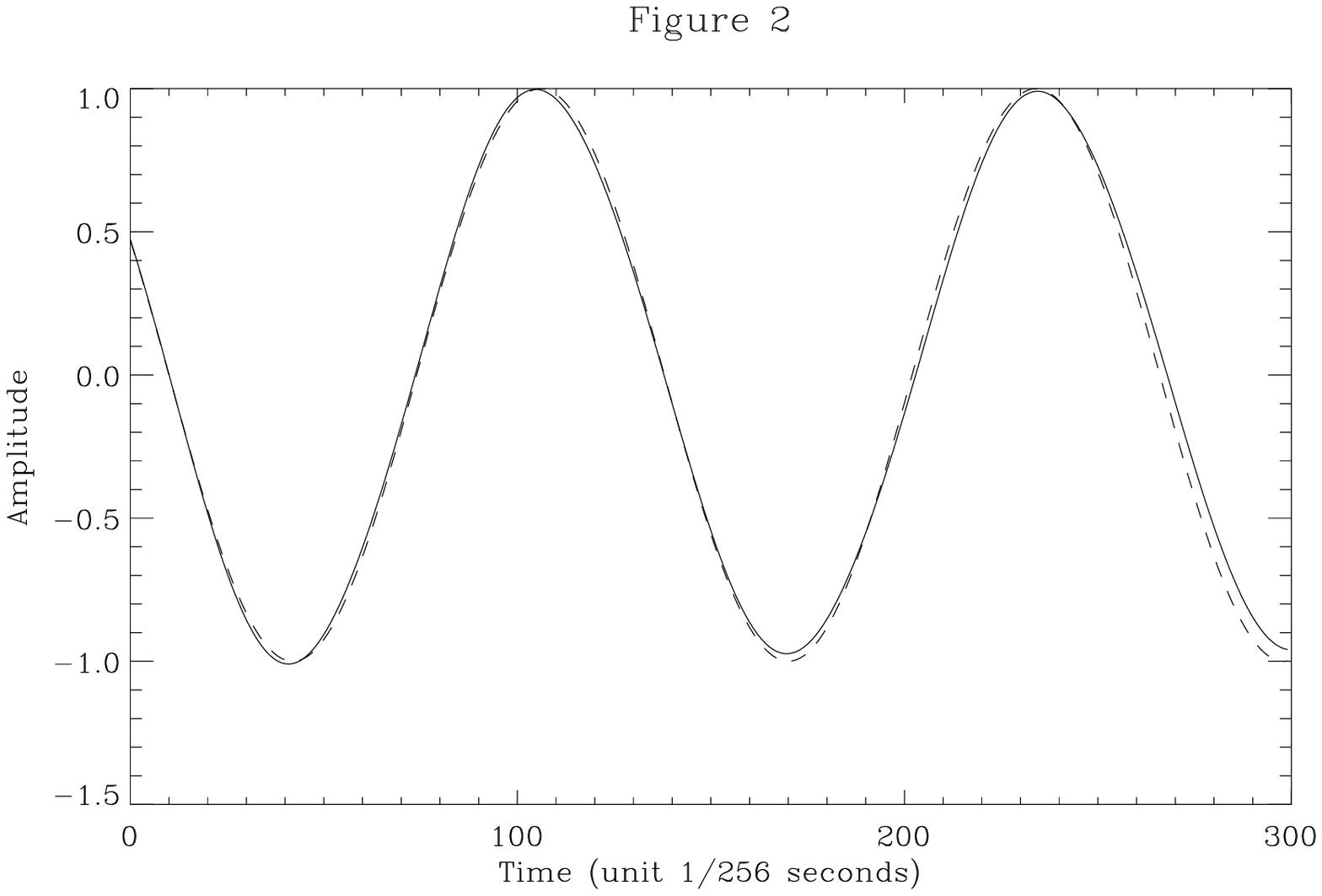}
\end{center}
\newpage
\begin{center}
\leavevmode
\epsfxsize 6in
\epsfbox{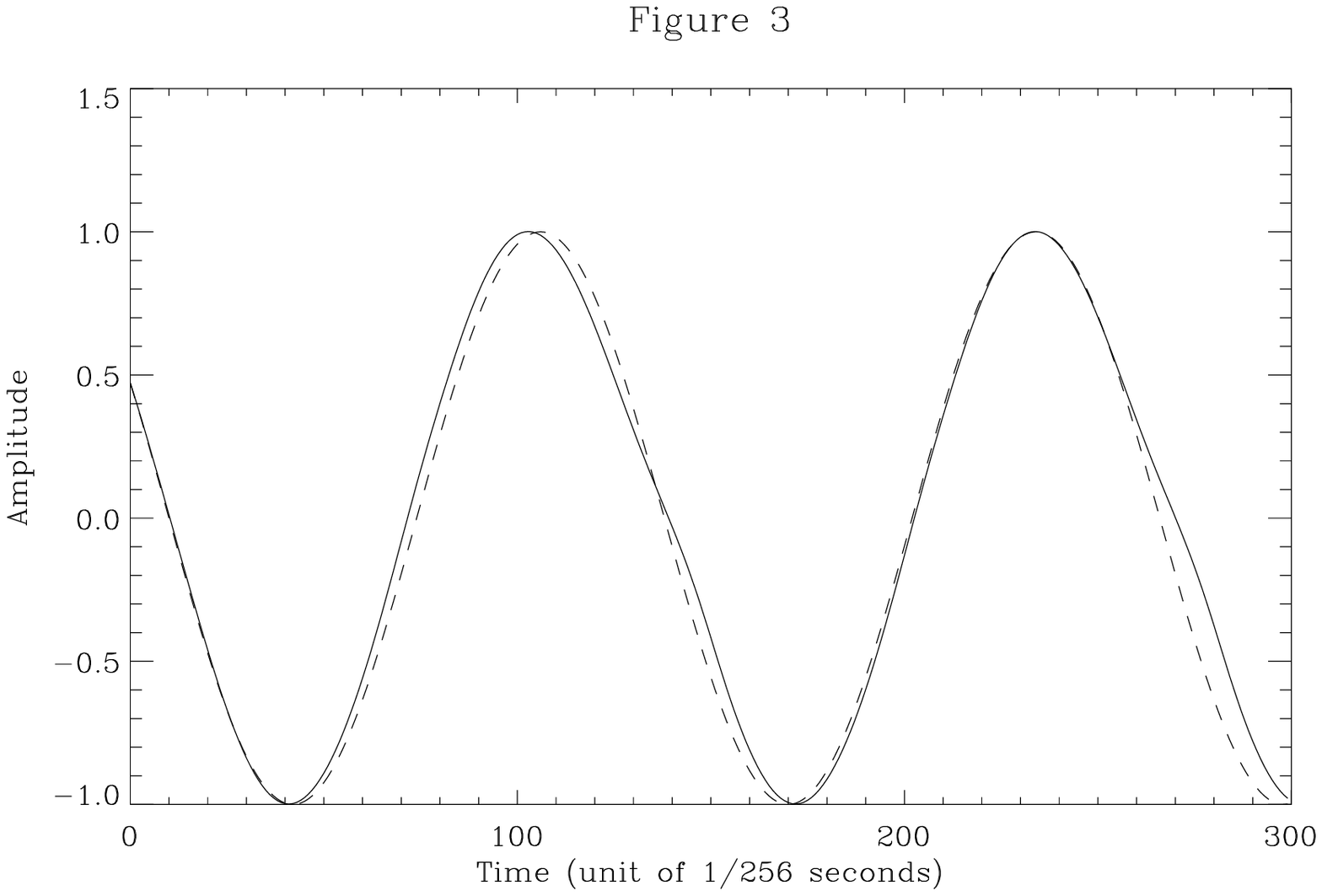}
\end{center}
\newpage
\begin{center}
\leavevmode
\epsfxsize 6in
\epsfbox{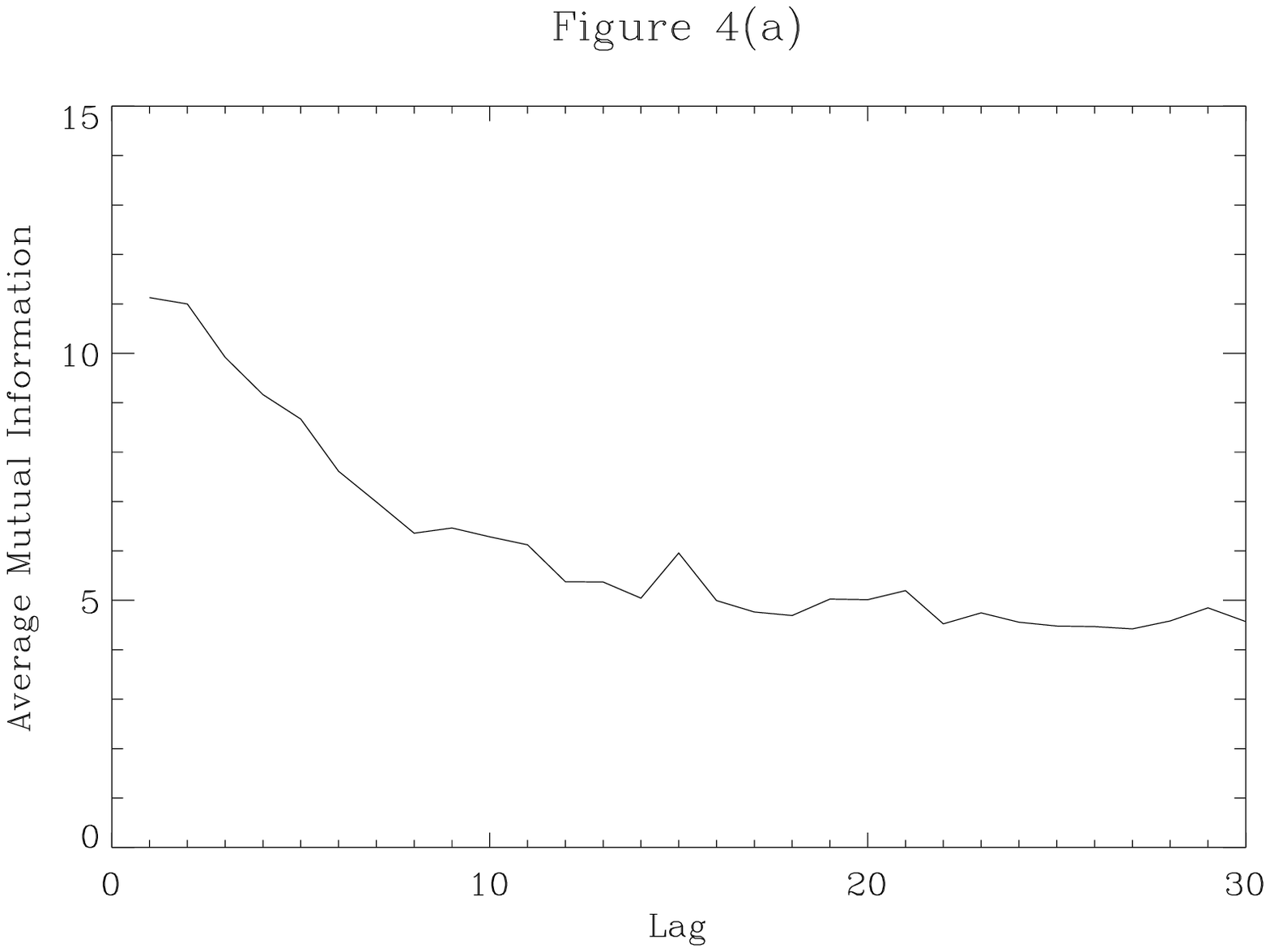}
\end{center}
\newpage
\begin{center}
\leavevmode
\epsfxsize 6in
\epsfbox{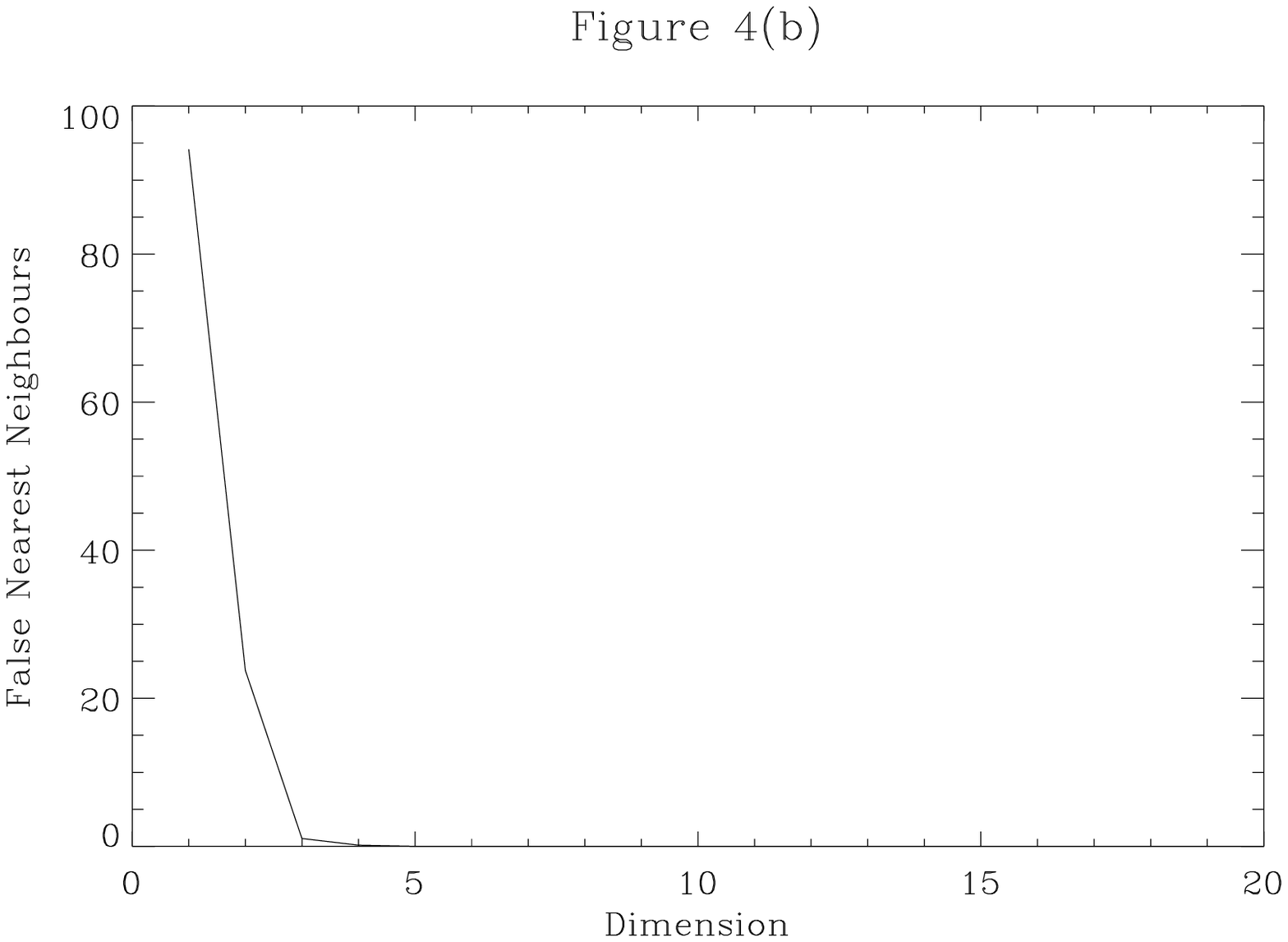}
\end{center}
\newpage
\begin{center}
\leavevmode
\epsfxsize 6in
\epsfbox{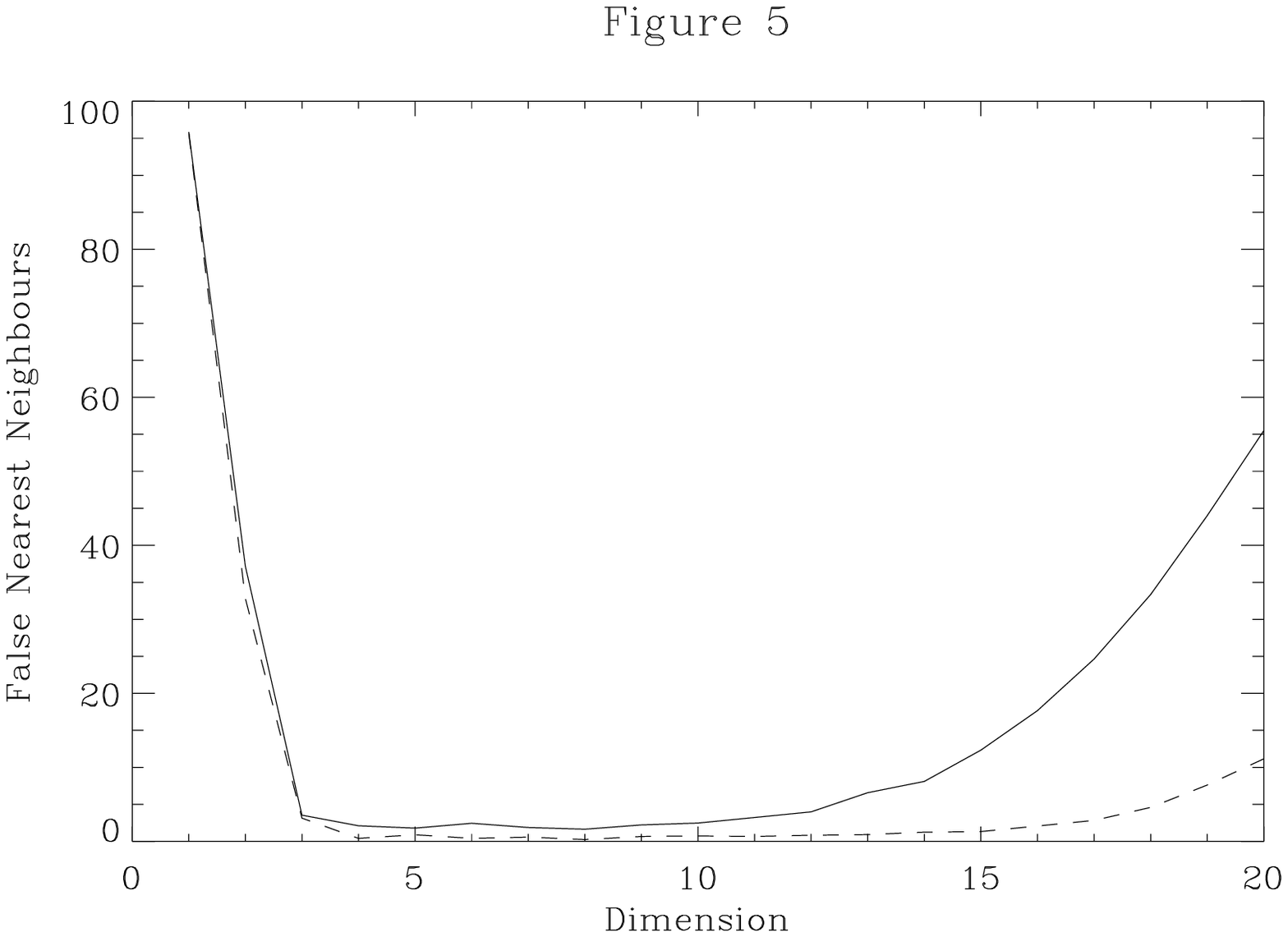}
\end{center}
\newpage
\begin{center}
\leavevmode
\epsfxsize 6in
\epsfbox{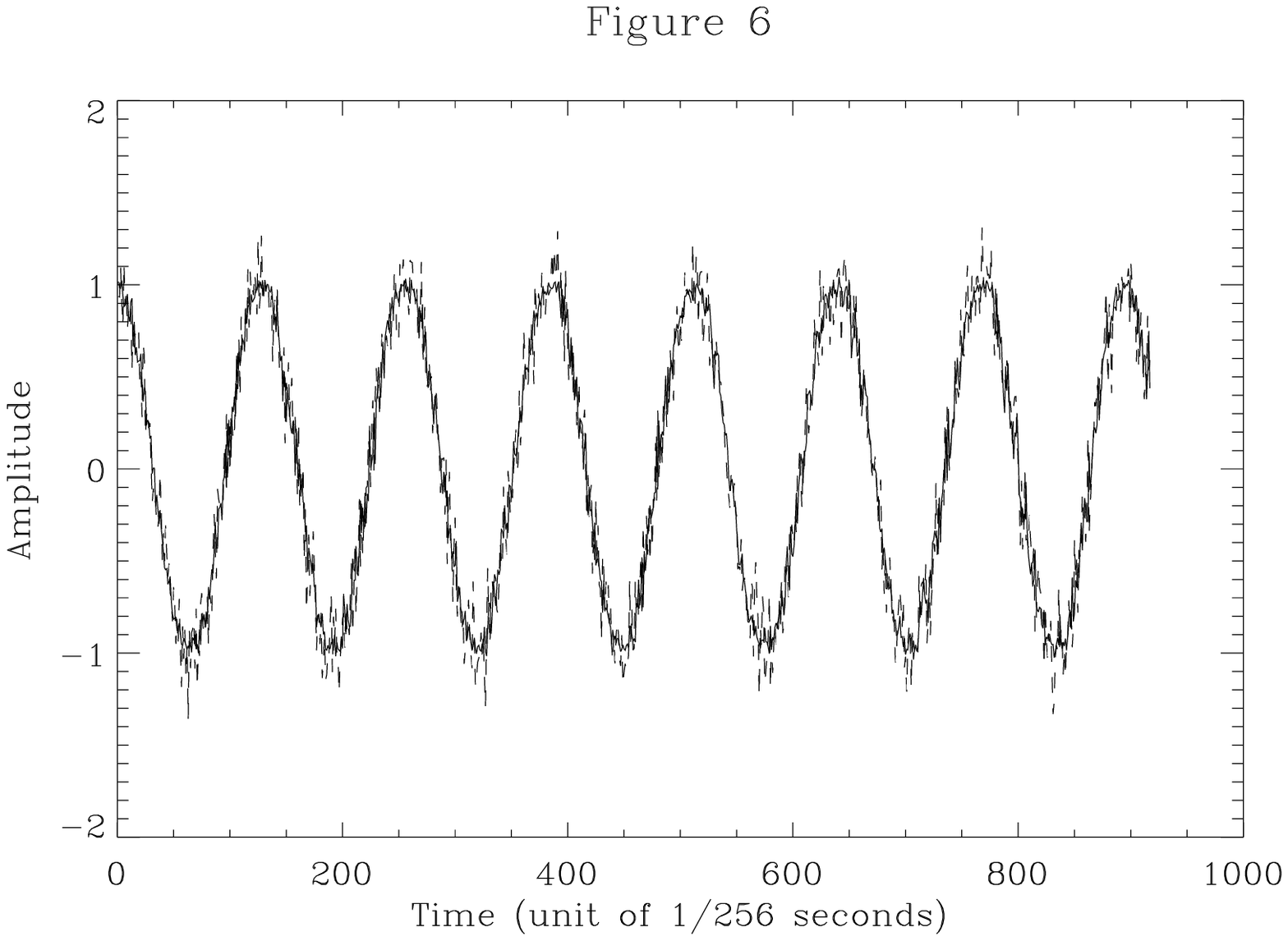}
\end{center}
\newpage
\begin{center}
\leavevmode
\epsfxsize 6in
\epsfbox{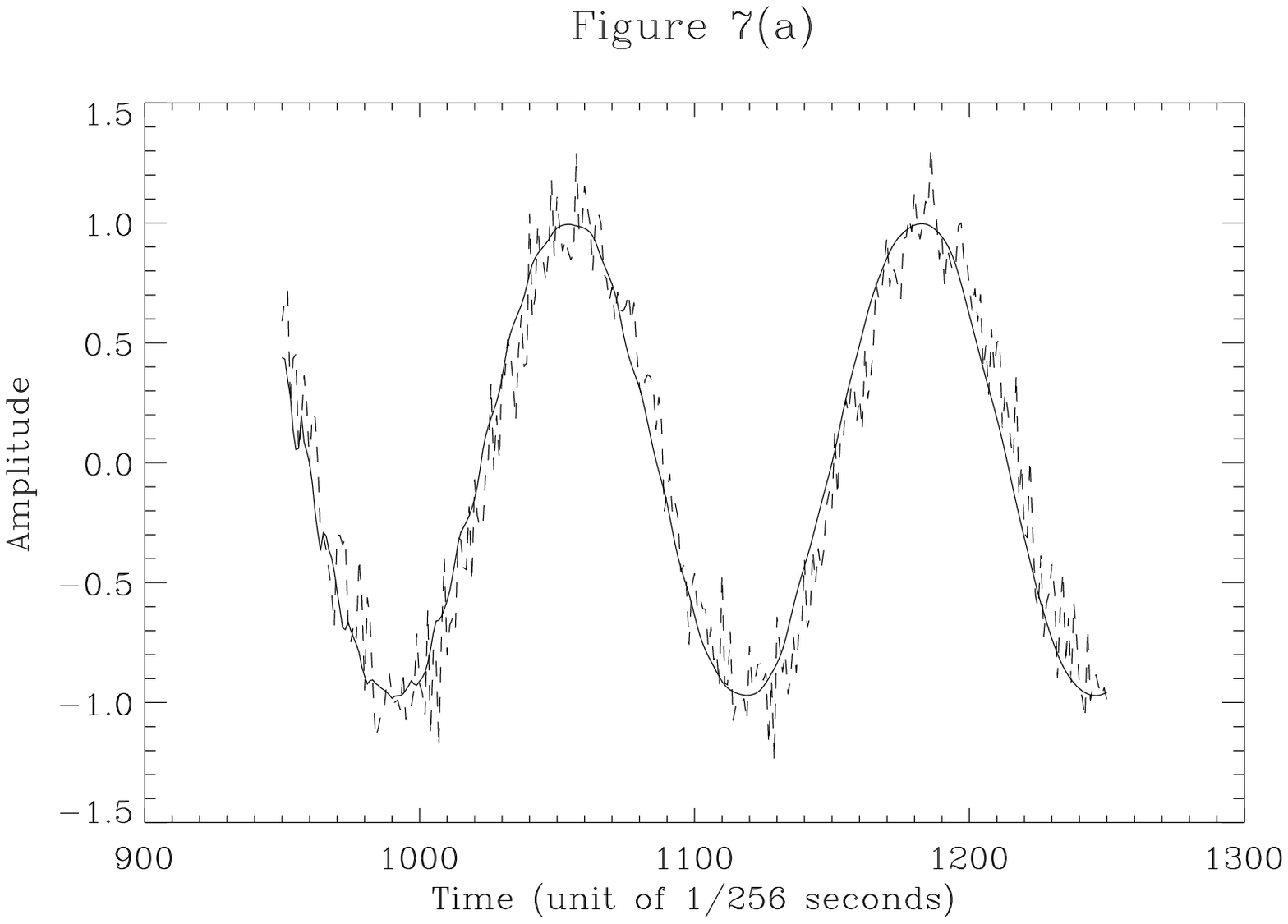}
\end{center}
\newpage
\begin{center}
\leavevmode
\epsfxsize 6in
\epsfbox{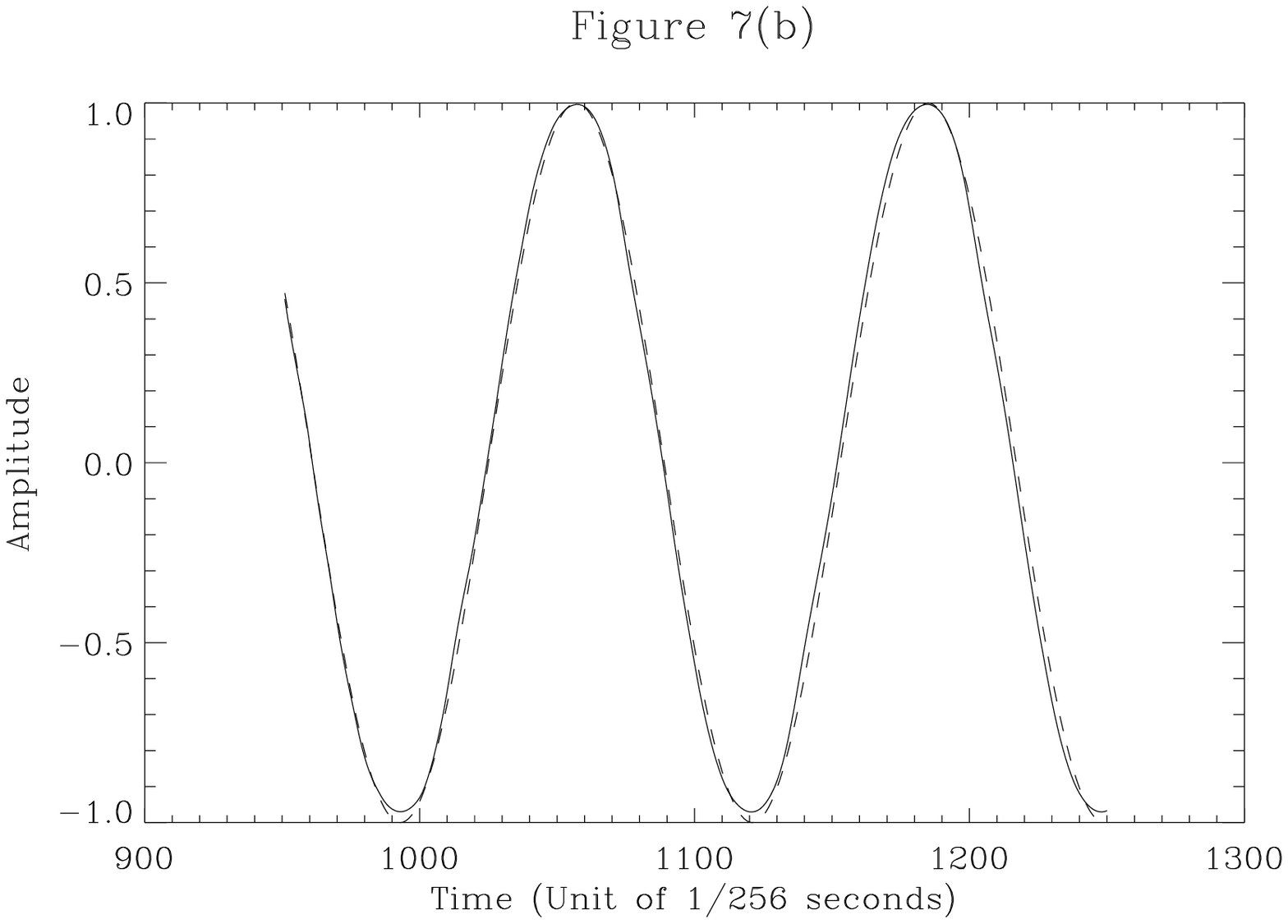}
\end{center}
\newpage
\begin{center}
\leavevmode
\epsfxsize 6in
\epsfbox{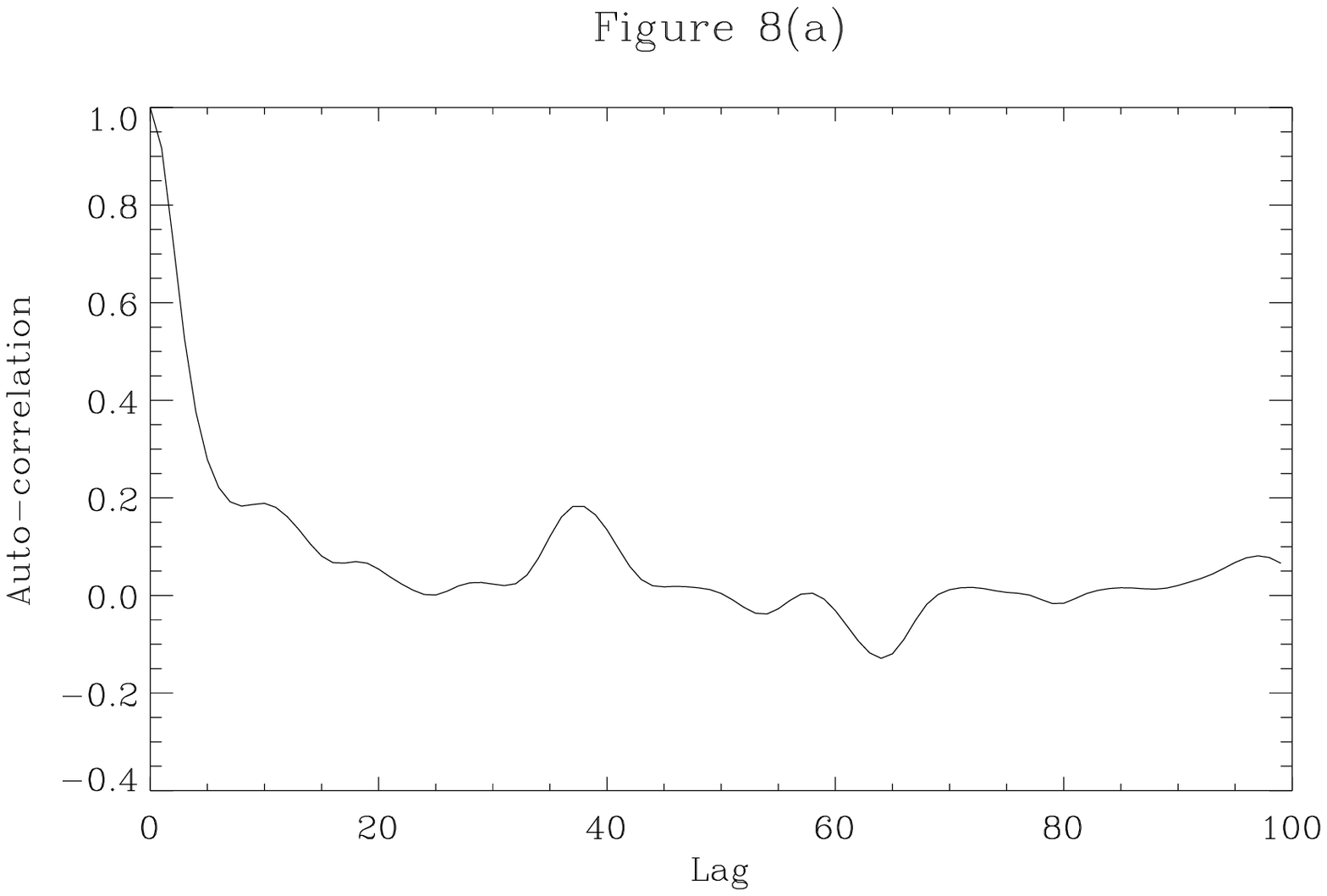}
\end{center}
\newpage
\begin{center}
\leavevmode
\epsfxsize 6in
\epsfbox{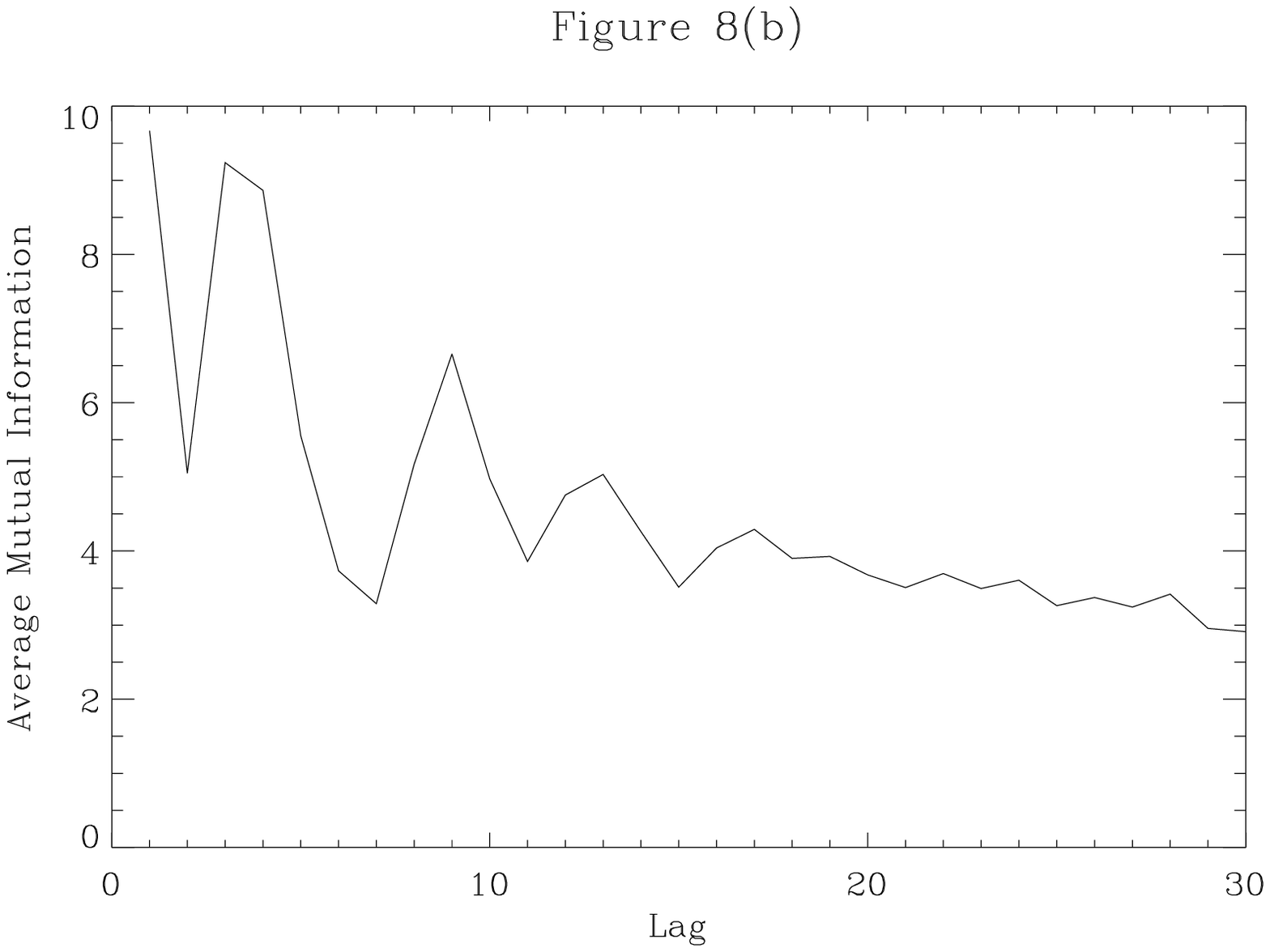}
\end{center}
\newpage
\begin{center}
\leavevmode
\epsfxsize 6in
\epsfbox{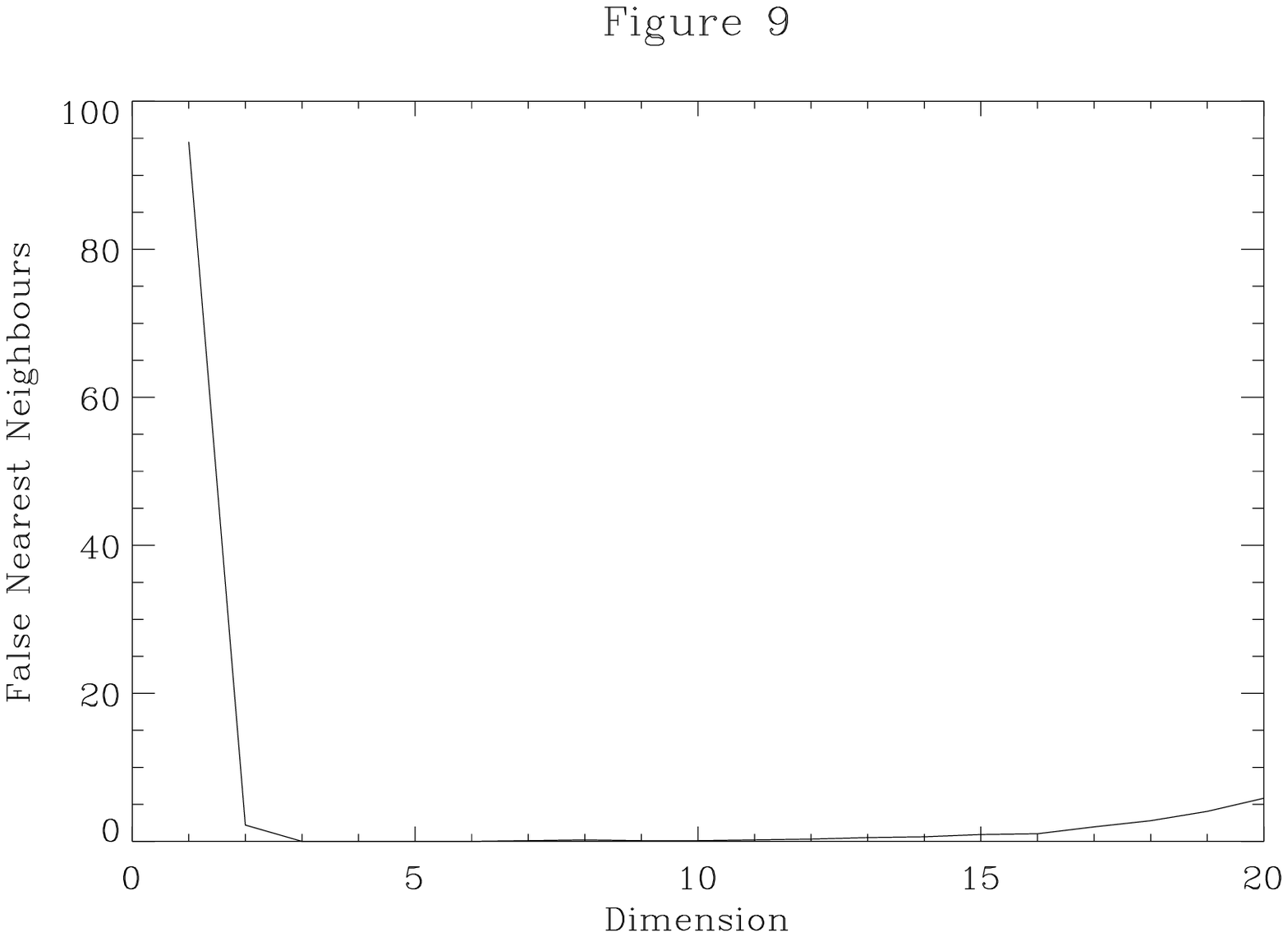}
\end{center}
\newpage
\begin{center}
\leavevmode
\epsfxsize 6in
\epsfbox{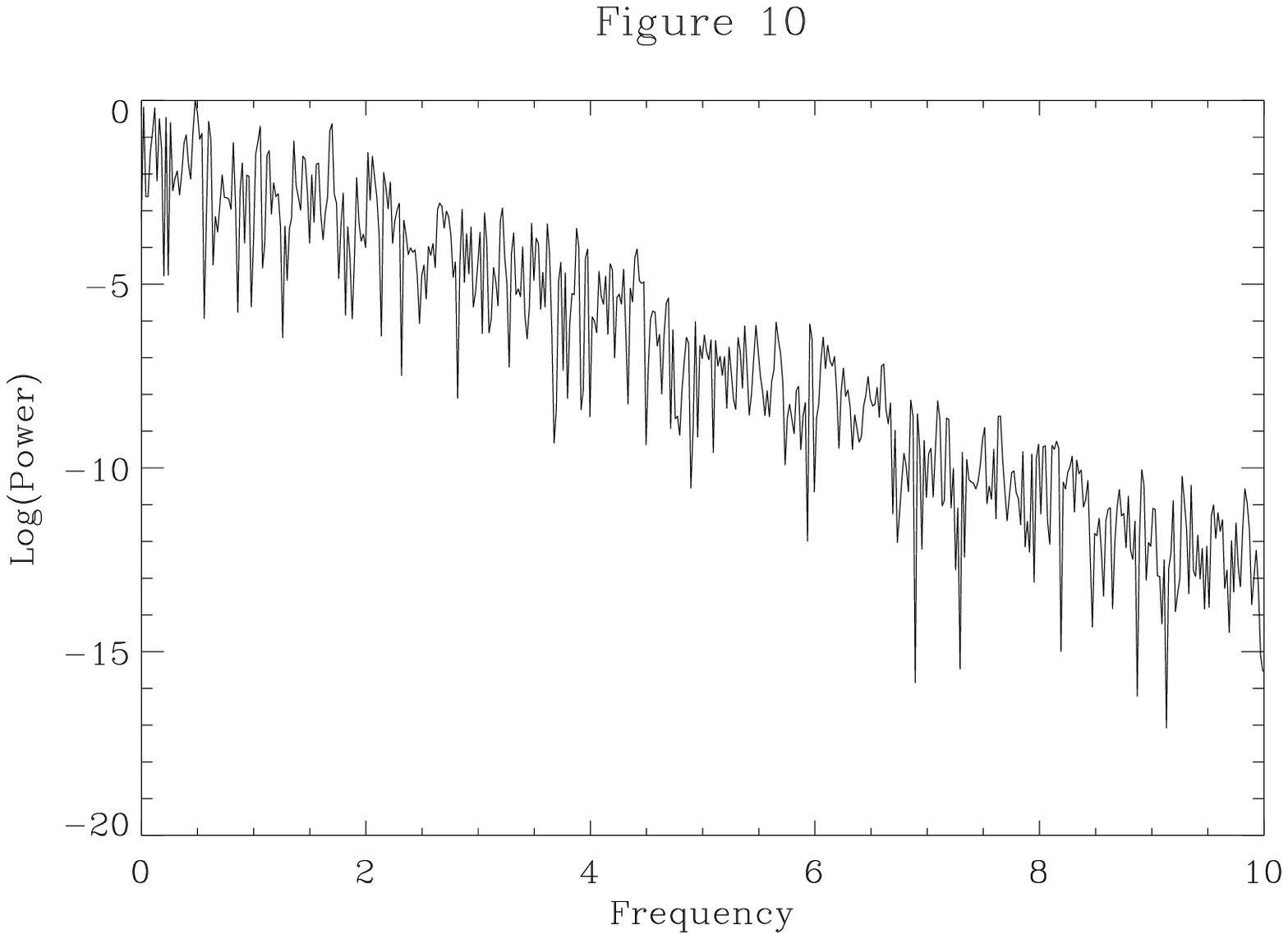}
\end{center}
\newpage
\begin{center}
\leavevmode
\epsfxsize 6in
\epsfbox{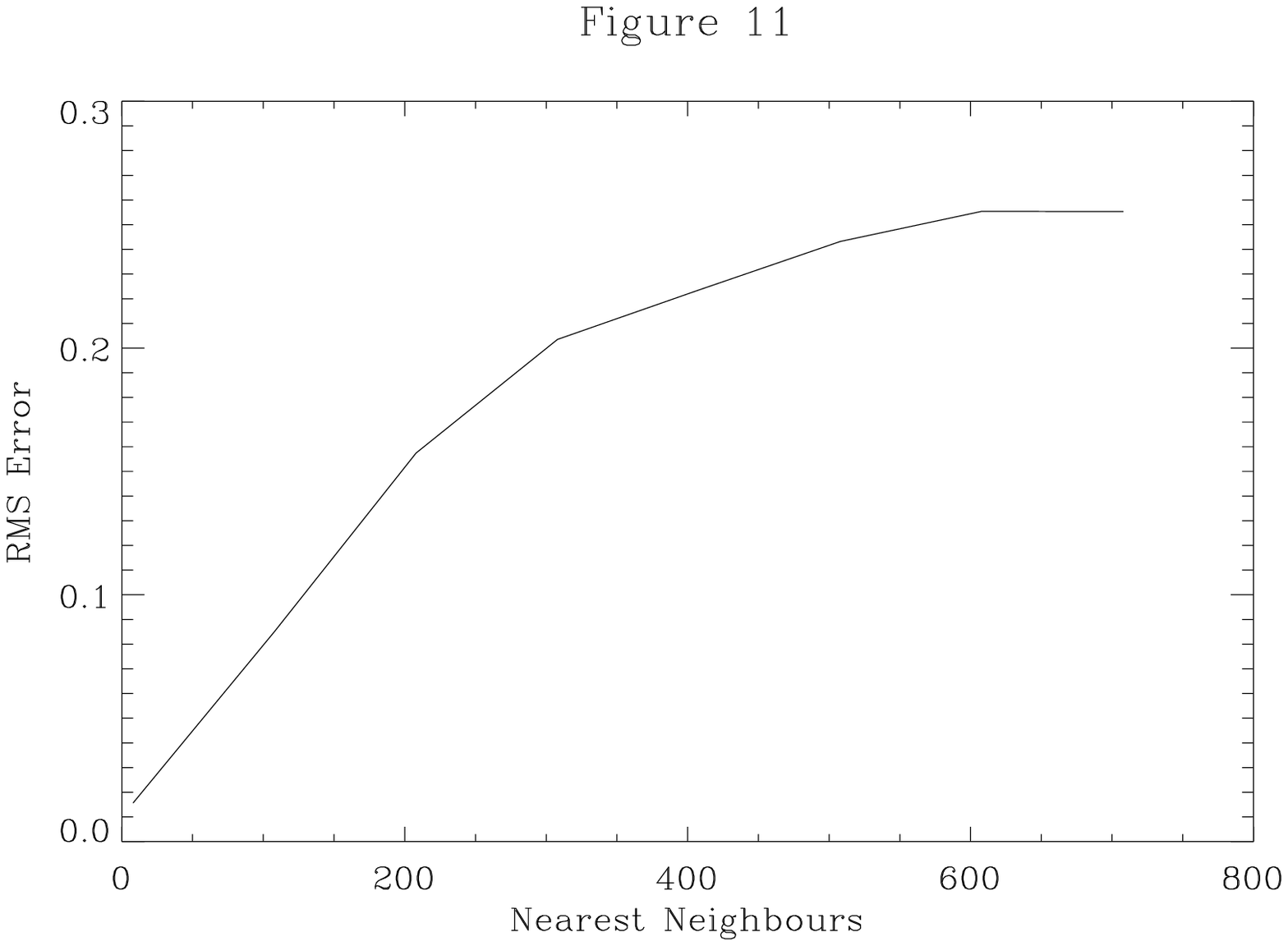}
\end{center}
\newpage
\begin{center}
\leavevmode
\epsfxsize 6in
\epsfbox{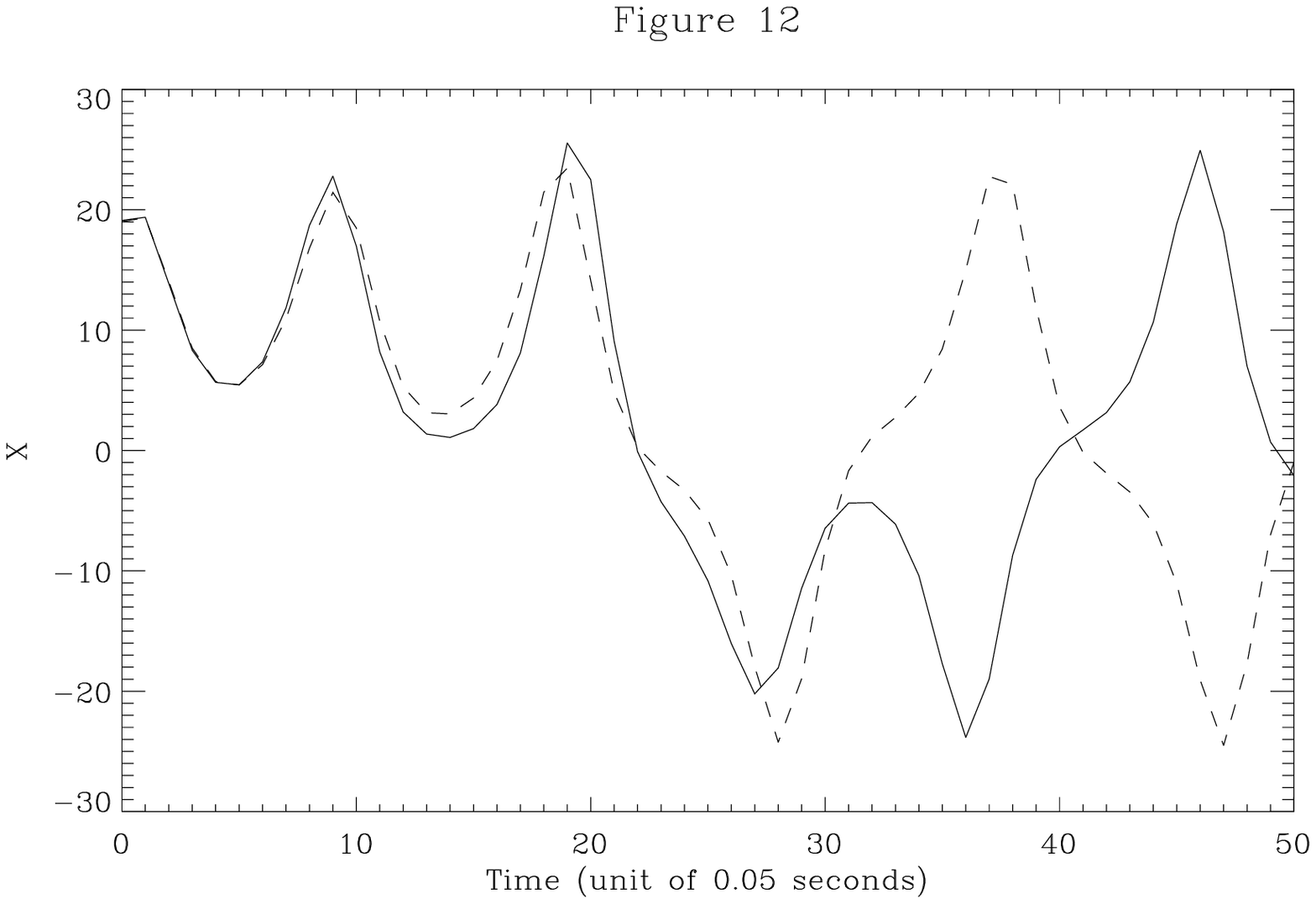}
\end{center}
\newpage
\begin{center}
\leavevmode
\epsfxsize 6in
\epsfbox{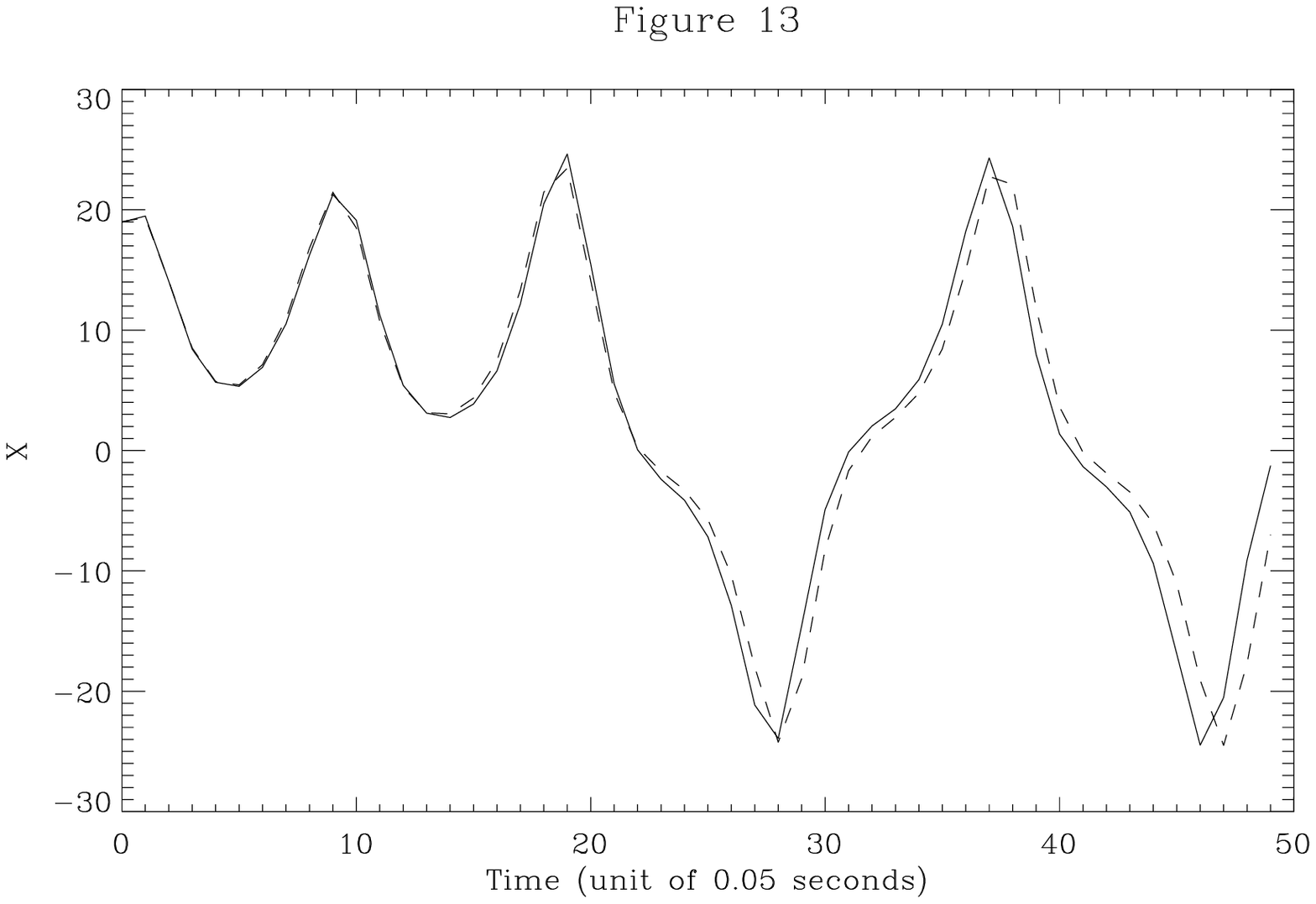}
\end{center}
\newpage
\begin{center}
\leavevmode
\epsfxsize 6in
\epsfbox{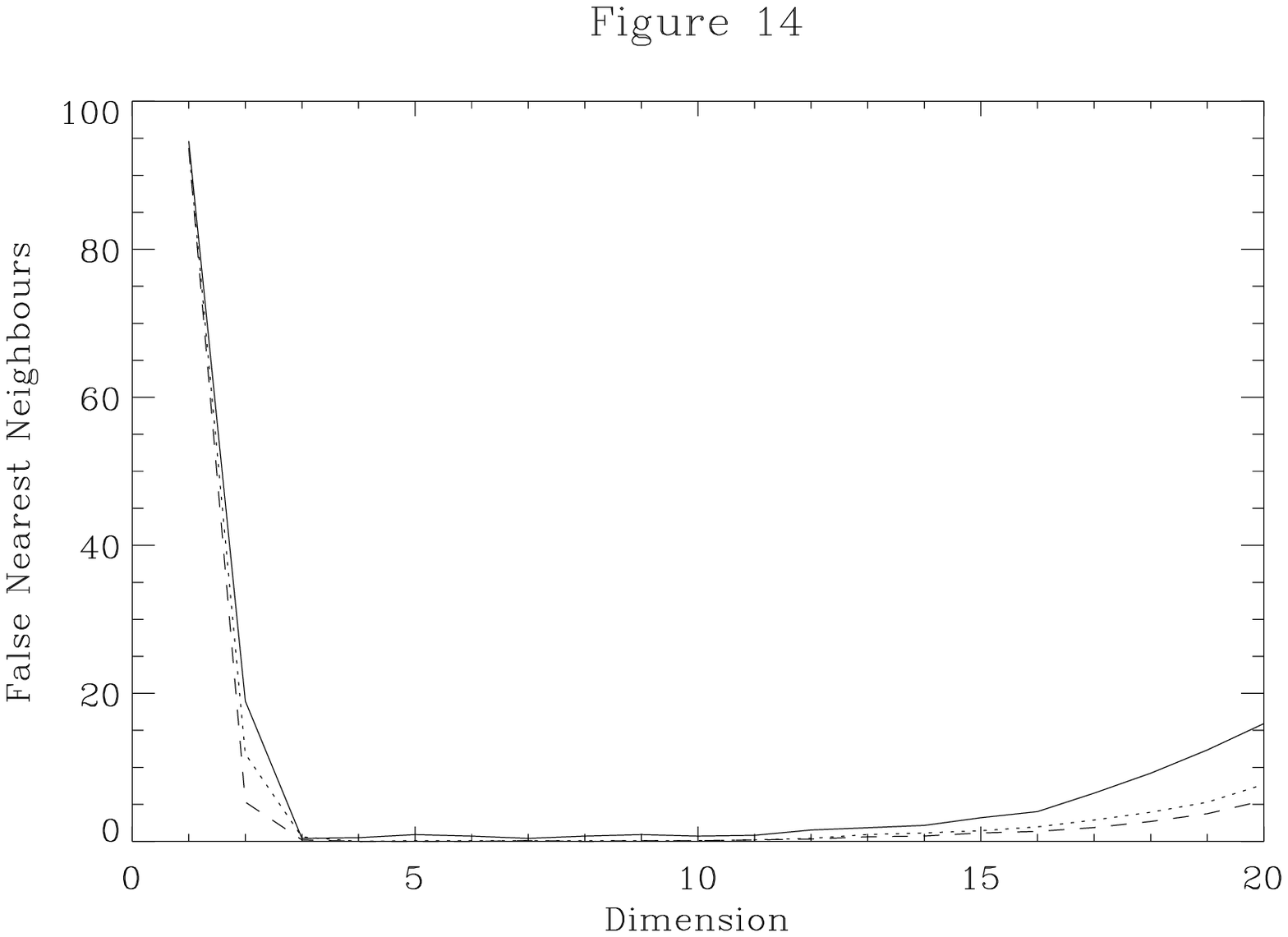}
\end{center}
\newpage
\begin{center}
\leavevmode
\epsfxsize 6in
\epsfbox{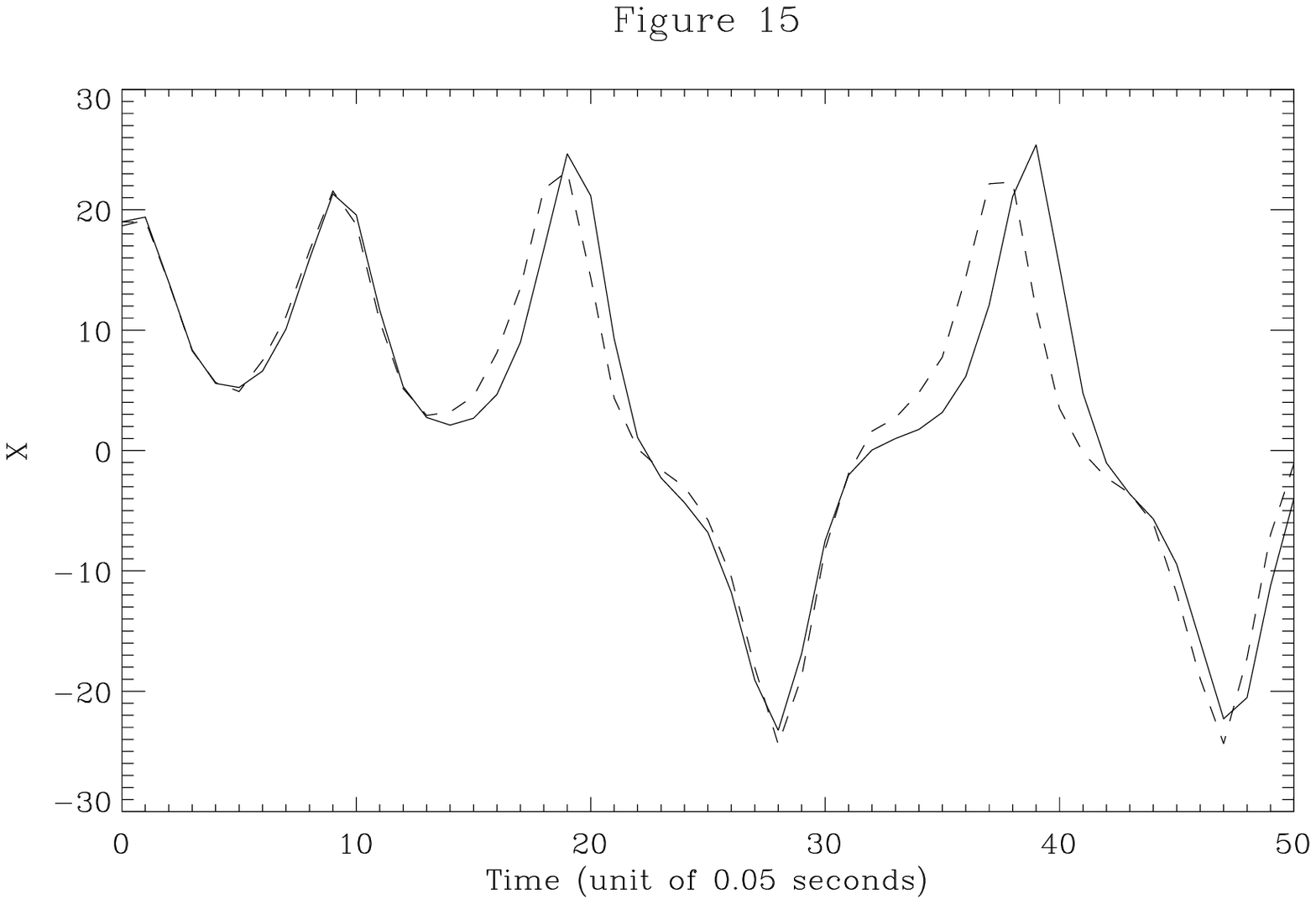}
\end{center}
\end{document}